\documentclass[aps,showpacs,amsmath,amssymb,lengthcheck]{revtex4}

\usepackage{epsfig}
\usepackage{color}
\usepackage{hyperref}

\begin{document}

\title{Critical wetting, first-order wetting and prewetting phase transitions\\ in binary mixtures of
Bose-Einstein condensates}

\author{{B. Van Schaeybroeck}}
\address{{Royal Meteorological Institute, Ringlaan 3, 1180 Brussels, Belgium.}}
\author{J.O. Indekeu}
\address{{Instituut voor Theoretische Fysica, Celestijnenlaan 200 D, KU Leuven, 3001 Leuven, Belgium.}}

\date{\small\it \today}

\begin{abstract}
An ultralow-temperature binary mixture of Bose-Einstein condensates adsorbed at an optical wall can undergo a wetting phase transition in which one of the species excludes the other from contact with the wall. 
Interestingly, while hard-wall boundary conditions entail the wetting transition to be of first order, using Gross-Pitaevskii theory we show that {\em first-order wetting as well as critical wetting} can occur when a realistic exponential optical wall potential (evanescent wave) with a finite turn-on length $\lambda$ is assumed. The relevant surface excess energies are computed in an expansion in $\lambda/\xi_i$, where $\xi_i$ is the healing length of condensate $i$. Experimentally, the wetting transition may best be approached by varying the interspecies scattering length $a_{12}$ using Feshbach resonances. In the hard-wall limit, $\lambda \rightarrow 0$, exact results are derived for the prewetting and first-order wetting phase boundaries. 
\end{abstract}

\pacs{03.75.Hh, 68.03.Cd, 68.08.Bc}

\maketitle

\section{Introduction and purpose}

\begin{figure}
\begin{center}
	\includegraphics[width=0.45\textwidth]{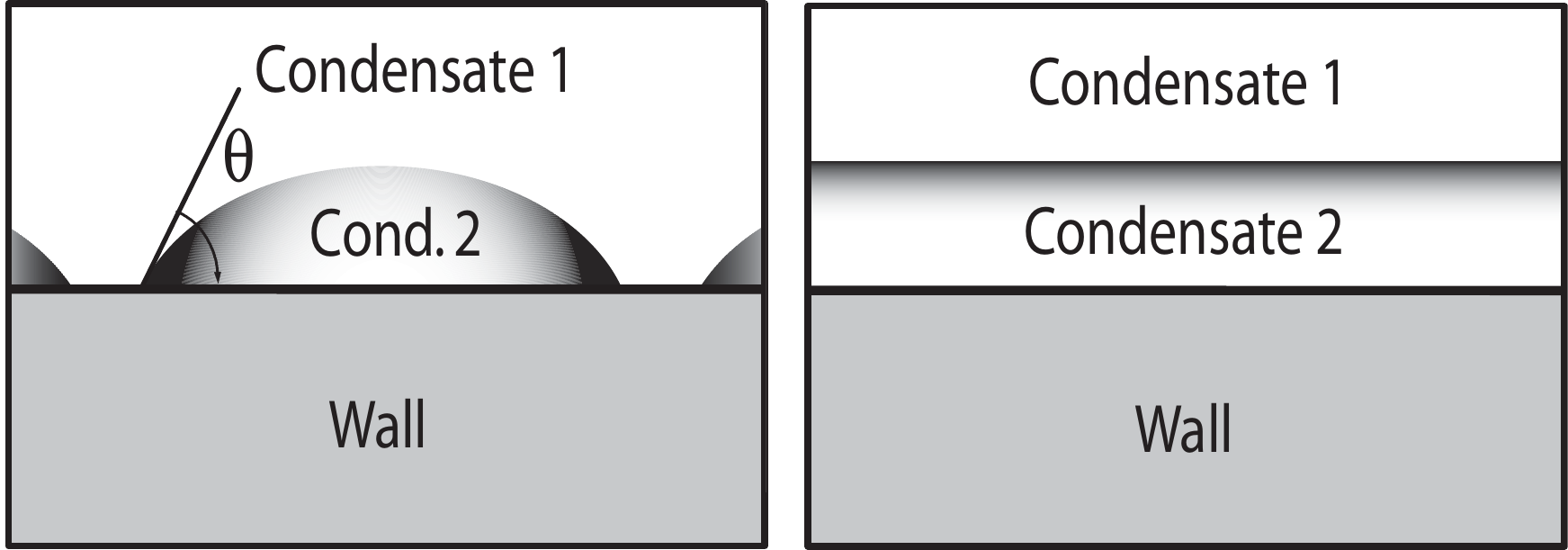}
	
       \caption{Left: Partial wetting. The interface between the two phases consisting of pure Bose-Einstein condensates 1 and 2 makes
       a finite contact angle $\theta$ with the optical wall. Right: Complete wetting. A macroscopic layer of pure phase
       $2$
       intrudes between the optical wall and pure phase $1$. \label{fig0ab}}
  \end{center}
\end{figure}

In a previous Letter \cite{indekeu} the possibility of wetting phase transitions \cite{degennes,Dietrich,BonnRMP} in mixtures of Bose-Einstein condensates (BECs) adsorbed at an optical wall was predicted based on Gross-Pitaevskii (GP) mean-field theory, in the limit of zero temperature ($T=0$). In a wetting phase transition, illustrated in Fig.1, a partial wetting state characterized by a thermodynamic contact angle or Young-Laplace angle $\theta$ undergoes a qualitative change in the limit $\theta \rightarrow 0$. In that limit a macroscopic layer of one of the two adsorbed phases, called the wetting phase, intrudes between the other phase and the wall, leading to complete wetting.

If we denote the excess (free) energy per unit area of the contact of condensate 1 (2) with the wall by $\gamma_{_{W1}}$ ($\gamma_{_{W2}}$) and the interfacial tension between condensates 1 and 2 by $\gamma_{_{12}}$, Young's law of mechanical equilibrium of a three-phase contact line reads \cite{RowlinsonWidom}
\begin{equation}
\gamma_{_{W1}} = \gamma_{_{W2}} + \gamma_{_{12}}\cos\theta,
\end{equation}
where $\theta$ is the thermodynamic contact angle (Fig.\ref{fig0ab}). Let us assume that condensate 2 has a lower surface energy than condensate 1, i.e., $\gamma_{_{W2}} < \gamma_{_{W1}}$. In this case we ask to what extent condensate 2 ``wets" the wall. The condition for partial wetting (PW) then reads
\begin{equation}\label{PW}
\gamma_{_{W1}} < \gamma_{_{W2}} + \gamma_{_{12}},
\end{equation}
and that for complete wetting (CW), also called Antonov's rule, is given by  (after thermodynamic equilibrium has been reached)
\begin{equation}\label{CW}
\gamma_{_{W1}} = \gamma_{_{W2}} + \gamma_{_{12}}.
\end{equation}
A wetting transition may occur in which $\theta \rightarrow 0$, i.e., a surface phase transition from PW, for which~\eqref{PW} holds, to CW, for which~\eqref{CW} is valid.

Conversely, in case $\gamma_{_{W1}} < \gamma_{_{W2}}$, the roles of the condensates are interchanged, and we ask to what extent condensate 1 ``dries" the wall. This change of terminology from wetting to drying is purely a matter of convention. It is inspired by a situation in adsorbed classical fluids, in which fluid 2 is a liquid and fluid 1 its vapour. In our BEC mixture, there is no physical distinction between wetting and drying. The terms merely alert us to the fact that for $\theta > 90^{\circ} $ the physical roles of labels 1 and 2 are interchanged.

When studying the wetting transition using Young's law a major simplification in the calculations can be implemented. The three relevant surface energies can be calculated using a one-dimensional geometry, such as in Fig.1 (Right), which is translationally invariant in both directions parallel to the wall. Knowledge of the surface energies allows one to deduce the contact angle through Young's law, without having to realize a two-dimensional inhomogeneity, such as in Fig.1 (Left). The two-dimensional problem depicted in Fig.1 (Left) can also be studied, for example, by applying an interface displacement model \cite{BonnRMP, indekeuFPSP}. This would allow one to obtain the structure of the so-called three-phase contact line, and its tension (energy per unit length), but this is outside the scope of the present work.

It is also worth mentioning that Young's law applies to every situation in which three phases are in mechanical equilibrium and in which surface excess energies ($T=0$) or surface excess free energies ($T>0$) can be defined, regardless of the nature of the microscopic forces. Young's law has been applied successfully not only to soft condensed matter systems, but also to hard condensed matter including ferromagnets and superconductors \cite{Dietrich, BonnRMP, indekeuFPSP}. In all these applications one has to keep in mind that $\theta$ is a ``thermodynamic" angle defined on a macroscopic scale \cite{RowlinsonWidom}.

The wetting transition in adsorbed BEC binary mixtures was shown to be of first order, with a discontinuity in the first derivative of the grand potential at the transition \cite{indekeu}. A number of extraordinary features emerged: (i) The grand potential is degenerate at wetting so that wetting layers of arbitrary thickness all have the same energy on the wetting phase boundary, (ii) the prewetting transition, being the continuation of the first-order wetting transition off of bulk two-phase coexistence, is critical (of second order), and corresponds to the nucleation of an infinitesimal prewetting film, whereas the prewetting transition is normally expected to be of first order, at least close to the first-order wetting point; and (iii) the prewetting line does not meet the bulk coexistence line tangentially at the wetting point, but under a finite angle, also at variance with expectations but nevertheless consistent with thermodynamics \cite{indekeu}. 

Experimental verification of this wetting transition was called for, especially in view of the fact that all the main physical parameters of the problem can be accurately controlled by applying an optical hard wall combined with a conventional harmonic trap to confine the particles to a half space and by applying an additional magnetic field to tune the interparticle forces through Feshbach resonances \cite{FeshbachReso}. Besides the report presented in \cite{indekeu}, a pedagogical discussion of these findings can be found in \cite{indekeuFPSP}.

Our main purpose in this paper is to show that the GP theory, for $T=0$, predicts that {\em the character of the wetting transition can change from first-order to critical} when the hard wall boundary condition is relaxed to a softer confining potential. In experiments this can be done using an  exponential wall potential, with a turn-on length $\lambda$ that is larger than the microscopic scattering length $a$ (typically 5 to 10 nm) but smaller than, or at most comparable to, the typical length scale associated with the spatial variation of the density profile, being the healing length $\xi$. The assumption of a hard wall has been a reasonable starting point for describing a set up with a surface trap, corresponding to an evanescent electromagnetic wave emerging from a prism. However, an exponential wall potential represents the optical wall  more realistically than a hard wall. The turn-on length $\lambda$ of the exponential is (at most) of the order of the wavelength of visible light divided by $4\pi$. In practice, this amounts to $\lambda \approx 50 $ nm. It is important to assess whether this length is still small compared to the two lengths that are relevant in the GP density-functional theory of BECs in a trap. Compared to the characteristic harmonic-oscillator length associated with the magnetic trap, $L$, which is of the order of 5 $\mu m$ or more, the turn-on length of the optical wall is small. Compared to the healing length $\xi$, which is the characteristic width of surface or interface inhomogeneities in the condensate fraction, and which is typically 200 to 400 nm, the length $\lambda$ is, however, not negligible. Therefore, it is important to refine the previous calculations by allowing for a softer wall. In sum, the length scales of our problem typically satisfy the following inequalities
\begin{equation}
a \ll \lambda \lesssim \xi \ll L
\end{equation}

The results we present are partly based on unpublished work \cite{BVSPhD} and make use of analytical calculations of the interfacial tension between two condensates and an exact expression for the first-order wetting phase boundary in the hard-wall limit \cite{BVS1}. The paper is organized as follows. In section II we recall the mean-field Gross-Pitaevskii description of the spatially varying condensate order parameters. Section III deals with the stability of bulk phases as a function of chemical potential and interaction strength. The excess grand potentials per unit area associated with the wall tensions and the interfacial tension are defined in section IV. 
Section V is devoted to the derivation of the phase diagram for nucleation, wetting and prewetting for the case of a hard-wall boundary condition. 
Our main new results are presented in section VI, which treats the wetting transitions encountered when the hard wall is replaced with a more realistic softer wall. Section VII treats the experimental relevance of our expressions for the surface tensions and our results for the wetting transitions. Some aspects of the presence of a harmonic trap are discussed in section VIII and section IX closes the paper with a conclusion and outlook.

\section{Mean-field theory for BEC binary mixtures}\label{sec_mixt}

When attempting to realize macroscopically phase-segregated phases in
Bose-Einstein systems, one tends to consider first the possibility of phase separation
between a (partially) condensed and a fully non-condensed state of a
single Bose gas. This, however, does not exist for ideal
gases~\cite{lamb,ziff}, nor does it exist for weakly interacting
ones due to the absence of a coexistence point between a
condensed and a fully normal phase. Therefore, spatial segregation
is only possible through the application of an external
potential~\cite{lamb,ziff}. In view of this, our attention shifted \cite{indekeu} to the investigation of possibilities for phase separation in {\em binary mixtures} of BECs. 

Since the experimental observation of weakly phase-segregated binary Bose-Einstein systems at the beginning of this century~\cite{modugno,miesner,myatt,stamper2,hall,matthews}, strong phase separation has been realized  more recently by at least six research groups~\cite{mccarron,tojo,altin,papp2,xiong} and even in a thermal mixture~\cite{baumer}, while many more degenerate Bose mixtures were produced in which phase separation is possible~\cite{stellmer,pilch,thalhammer}. The physics of multi-component condensates is well explained in Refs.~\cite{stamper3,malomed2} both focusing on theory and experiments. While the statics and dynamics of phase-separated BECs have been extensively studied in Refs.~\cite{ho,ejnisman,pu,alexandrov,timmermans,ao,svidzinsky,svidzinsky2,navarro,wen,ronen,pattinson,gautam}, the phenomenology associated with the interface in Bose mixtures was explored in Refs.~\cite{bezett,sasaki,ticknor,takeuchi,goldman,kadokura,pepe} and the phase diagram at finite temperature was investigated in Refs.~\cite{phat,BVS2,shi}.

The Gross-Pitaevskii (GP) formalism provides us with a mean-field
equation of state for Bose gases at $T \approx 0$. It is generally used
for dilute, weakly-interacting gases at ultralow temperatures. In
the following, we consider two condensates present in a volume $V$
at chemical potentials $\mu_1$ and $\mu_2$, respectively. An
appropriate mean-field energy functional is found by applying a
Bogoliubov approximation~\cite{fetter,pitaevskii} that reduces the particle
field operators $\widehat{\psi}_i({\bf r})$ to a sum of their
ground state mean value and a fluctuation term:
$\widehat{\psi}_i({\bf r})=\psi_i({\bf
r})+\delta\widehat{\psi}_i({\bf r})$ ($i=1,\,2$). Here,
$|\psi_i({\bf r})|^2$ equals the (local) mean density $n_i({\bf r})$ of the Bose
condensed atoms of species $i$.  Due to the ultralow
temperature, the potential of the particle interactions can for calculational purposes be replaced with a contact potential (also called Fermi pseudo potential) 
$V_{ij}(\mathbf{r}-\mathbf{r}')=\delta(\mathbf{r}-\mathbf{r}')\,G_{ij}$.
Expanding the full second-quantized grand potential to zeroth
order in $\delta\widehat{\psi}_{i}$ ($i=1,2$), one
obtains~\cite{fetter}:
\begin{align}\label{vrije}
\Omega =&\sum_{i=1,2}\int_{V}\text{d}\mathbf{r}\left(
\psi_i^{*}(\mathbf{r})\left[-\frac{\hslash^{2}}{2m_{i}}
\boldsymbol{\nabla}^{2}-\mu_i+U_i(\mathbf{r})\right]
\psi_i(\mathbf{r})\right.\nonumber\\
&\left.+\frac{G_{ii}}{2}|\psi_i(\mathbf{r})|^{4}\right)+G_{12}
\int_{V}{\text{d}\mathbf{r}\,|\psi_1(\mathbf{r})|^{2}
|\psi_2(\mathbf{r})|^{2}},
\end{align}
where $U_i$ is the external trapping potential of species $i$ and
the coupling constants $G_{ij}$ are linear in the $s$-wave
scattering lengths $a_{ij}$ and depend on the particle masses through the
identity $G_{ij}=2\pi \hslash^{2}a_{ij}\left(1/m_i+1/m_{j}\right)$
with $i,\,j=1,\,2$. The use of fixed chemical potentials instead of
fixed particle numbers is justified since our semi-infinite system
can be viewed as an open system which is in direct contact with
``bulk reservoirs'' of condensate, so that the
number of atoms can change without
affecting the thermodynamical properties of the system as a whole.
Moreover, it can readily be checked that Young's law and consequently the
  phase diagrams for wetting at a hard wall are the same in the canonical ensemble (CE) and the grand canonical ensemble (GCE). Indeed, the surface excess energies and the interfacial tension defined in the GCE are equal to 4 times their counterparts in the CE \cite{BVS1,BVSPhD}. These counterparts are related but physically distinct quantities.

In the absence of particle flow, one chooses the order parameters to be real valued. Demanding the first variation of the energy functional to vanish leads us then to the coupled GP equations
\begin{subequations}\label{GP1}
\begin{align}
\frac{\hslash^{2}}{2m_1} \boldsymbol{\nabla}^{2}\psi_1=
(U_1-\mu_1)\psi_1+G_{11}\psi_1^{3}+G_{{12}}
\psi_2^{2}\psi_1,\\
\frac{\hslash^{2}}{2m_2} \boldsymbol{\nabla}^{2}\psi_2=
(U_2-\mu_2)\psi_2+G_{22}\psi_2^{3}+G_{12}\psi_1^{2}\psi_2.
\end{align}
\end{subequations}
The equilibrium pressure for a pure and homogeneous phase of species
$i$ with $U_i(\mathbf{r})=0$ is ($i=1,\,2$)
\begin{align}\label{drukken}
P_i=\left.-\frac{\partial \Omega_i}{\partial
V}\right|_{\psi_i^2=n_i}=\frac{\mu^{2}_i}{2G_{ii}},
\end{align}
where $\Omega_i$ is the grand potential of pure species $i$ and $n_i$ is its homogeneous density. If present as a pure phase, species $1$ has a density $n_1 = \overline{n}_{1}\equiv\mu_1/G_{11}$. Each value of $\mu_1$ can be associated with a value of the chemical potential for species $2$, defined by $\overline{\mu}_2\equiv\mu_1\sqrt{G_{22}/G_{11}}$, so that at two-phase coexistence (when $P_2=P_1$), $\mu_2=\overline{\mu}_2$.
Define then also the density $\overline{n}_2\equiv
\overline{\mu}_2/G_{22}$. We rescale now the order parameters
$\psi_1$ and $\psi_2$ and define the normalized wave functions
$\widetilde{\psi}_1$ and $\widetilde{\psi}_2$ and densities
$\widetilde{n}_1$ and $\widetilde{n}_2$:
\begin{subequations}
\begin{align}
\widetilde{\psi}_1&\equiv\psi_1/\sqrt{\overline{n}_1} = \sqrt{n_1/\overline{n}_1} \equiv \sqrt{\widetilde{n}_1},\\
\widetilde{\psi}_2&\equiv\psi_2/\sqrt{\overline{n}_2} = \sqrt{n_2/\overline{n}_2}\equiv
\sqrt{\widetilde{n}_2}.
\end{align}
\end{subequations}
Note that the normalization is with respect to the bulk densities of pure phase 1 and of pure phase 2 at coexistence with phase 1. This is mathematically convenient, but implies that while the scaled order parameter for a pure and homogeneous phase 1 equals 1, this is not the case for phase 2, except at two-phase coexistence. These definitions are convenient whenever (at least) phase 1 is stable in bulk, which we will always assume.

The quantum nature of the system results in zero-point motion;
this determines the typical length scale for density modulations,
and therefore also the thickness of surface inhomogeneities at the boundaries, the
vortex-core size and the size of soliton like structures in the interior. This
quantum effect is embodied in the gradient (or Laplacian) terms in the nonlinear
Schr\"{o}dinger equation~\eqref{GP1}. The resulting lengths are the
healing lengths $\xi_1$ and $\xi_2$ which are defined as:
\begin{align}\label{healing}
\xi_1=\frac{\hslash}{\sqrt{2m_1\mu_1}}\quad\text{and}\quad
\xi_2=\frac{\hslash}{\sqrt{2m_2\mu_2}}\,.
\end{align}
Again for our convenience, an auxiliary healing length $\overline{\xi}_2$ for species 2 is defined in terms of the chemical potential of species 2, when species 2 is at two-phase coexistence with species 1:
\begin{align}\label{healingoverline}
\overline{\xi}_2  = \frac{\hslash}{\sqrt{2m_2\overline{\mu}_2}} = \sqrt{\frac{\mu_2}{\overline{\mu}_2}}\;\xi_2.
\end{align}

The interparticle interaction strength $G_{12}$ is only
relevant in comparison with the geometric mean of $G_{11}$ and
$G_{22}$. Hence we define $K\equiv G_{12}/\sqrt{G_{11}G_{22}}$.
Both $K$ and $\overline{\xi}_2/\xi_1$ can be expressed in terms of the atomic
masses and the scattering lengths:
\begin{align}\label{scatter}
K=\frac{m_1+m_2}{2\sqrt{m_1 m_2}}\frac{a_{12}}{\sqrt{a_{11}a_{22}}}
\quad\text{
 and  }\quad \overline{\xi}_2/\xi_1=\sqrt[4]{\frac{m_1a_{11}}{m_2a_{22}}},
\end{align}
where also the latter relation does not assume two-phase coexistence, by virtue of how we defined $\xi_2 $ by means of $\overline{\mu}_2$ instead of $\mu_2$. Finally, after rescaling space $\mathbf{r}=\xi_1\widetilde{\mathbf{r}}$, the GP Eqs.~\eqref{GP1}
reduce to:
\begin{subequations}\label{GP2}
\begin{align}
 \boldsymbol{\nabla}^2\widetilde{\psi}_1=&
(U_1-\mu_1)\widetilde{\psi}_1/\mu_1+\widetilde{\psi}_1^{3}+K\widetilde{\psi}_2^{2}\widetilde{\psi}_1,\label{GP2a}\\
[\overline{\xi}_2/\xi_1]^2 \boldsymbol{\nabla}^2\widetilde{\psi}_{2}=&
(U_2-\mu_2)\widetilde{\psi}_2/\overline{\mu}_2+\widetilde{\psi}_2^{3}+K\widetilde{\psi}_1^{2}\widetilde{\psi}_2.\label{GP2b}
\end{align}
\end{subequations}
For the special case of hard wall boundary conditions, implying $U_i=0$ when $\psi_i\neq 0$, and assuming the presence of pure and homogeneous phase $1$ somewhere in the considered volume (e.g., far from the optical wall), the first integral of the GP equations is:
\begin{align}\label{behoudenergie}
( \boldsymbol{\nabla}\widetilde{\psi}_1)^{2}+[\overline{\xi}_2/\xi_1]^2(
\boldsymbol{\nabla}\widetilde{\psi}_2)^{2}+&
\widetilde{\psi}_1^{2}+[\mu_2/\overline{\mu}_2]\widetilde{\psi}_2^{2}\\
&-\frac{\widetilde{\psi}_1^{4}}{2}
-\frac{\widetilde{\psi}_2^{4}}{2}-K
\widetilde{\psi}_1^{2}\widetilde{\psi}_2^{2}=\frac{1}{2},\nonumber
\end{align}
where the constant 1/2 results from the observation that far from the wall, for $z \rightarrow \infty$, the order parameters reach their bulk values $\widetilde{\psi}_1 = 1$ and $\widetilde{\psi}_2 = 0$ (and their derivatives vanish). Equivalently, the constant can be evaluated at the hard wall boundary, at $z=0$, in the presence of condensate 1 alone. Then, the derivative of $\widetilde{\psi}_1$ at $z=0$ takes the value $1/\sqrt{2}$ and $\widetilde{\psi}_1(0)=0$.

\section{Thermodynamics of BEC Mixtures}\label{sec_thermo}
\begin{figure}
\begin{center}
	\includegraphics[width=0.45\textwidth]{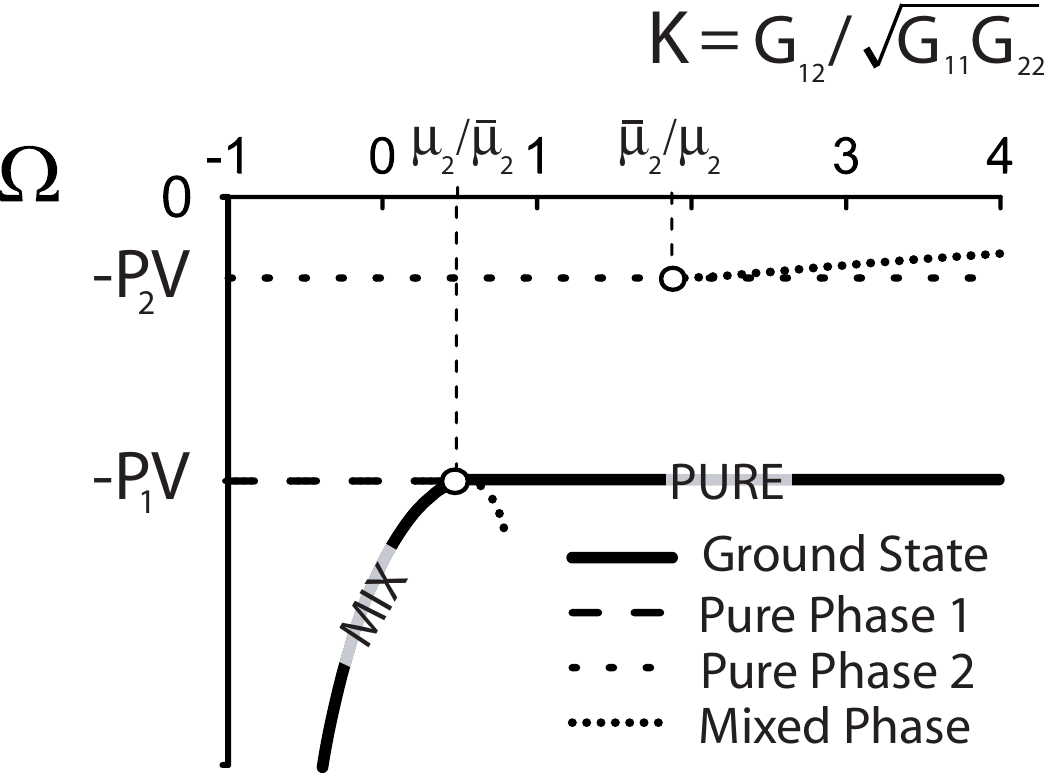}
       \caption{The grand potential of the three possible bulk states (and the vacuum at $\Omega=0$) as a function of the
       relative interaction parameter $K$ when $\mu_2<\overline{\mu}_2$. The ground-state grand potential is indicated by the thick line.
       For $-1<K\leq\mu_2/\overline{\mu}_2$, the ground state coincides with the mixed phase
       and for $\mu_2/\overline{\mu}_2<K$, it coincides with pure phase $1$. The first derivative of $\Omega$ is continuous in this transition (second-order phase
       transition), as is evidenced by the mathematical continuation of the energy of the mixed phase (curved dotted line).
       When $K\geq \overline{\mu}_2/\mu_2$, the mixed phase is an unstable state (ascending dotted line).\label{fig0a}}
  \end{center}
\end{figure}
Our presentation in this section introduces no new physics beyond what was found before (see especially \cite{ao}), but recapitulates the bulk properties of BEC mixtures in a way suitable for our further purpose.

Experimentally, the number of particles in a condensate is finite and fixed (neglecting losses) and this calls for a description in a CE. Theoretically, it is more practical to work in a GCE invoking hypothetical reservoirs or ``baths" at fixed chemical potential. The two descriptions are related. Both allow one to define quantities with the dimension of excess energy per unit area. The surface excess energy that we calculate grand canonically, differs by a factor of 4 from its Legendre-conjugated quantity in the CE~\cite{fetter,indekeu,BVSPhD}. Consequently, for the application of Young's law this factor of 4 drops out, and we are free to work with either definition.

We clarify now the equilibrium bulk ``phases'' found when putting two species in a volume $V$, in contact with two baths which are
at fixed chemical potentials: The ground states are either one of the two \textit{pure phases} or the \textit{mixed phase} where the latter has by definition  nonzero densities for both species.  The parameter $1/K$ turns out to play a role analogous to that of the temperature $T$ for ordinary liquid mixtures (with fixed microscopic interactions) in the sense that for a small value of $1/K$ (or temperature $T$ for ordinary, classical liquids) the species tend to demix whereas for a value larger than $1/K=1$ (or $T$ greater than some consolute temperature $T_{c}$ for classical binary liquid mixtures), they mix.

A volume $V$ containing pure and homogeneous phase $i$ has a grand potential
$\Omega_i=-P_iV$ where the pressure $P_i$ is given by~\eqref{drukken}. Since the grand potential of the
vacuum is zero, the pure phase can constitute the ground state
whenever $G_{ii}>0$ (repulsive interactions). Now assume for a moment that $P_1\geq P_2$.
When the relative interaction parameter $K$ satisfies
$-1<K\leq\mu_2/\overline{\mu}_2 = \sqrt{P_2/P_1}$, the mixed phase ($M$) minimizes the
grand potential (i.e.,~$\Omega_M<\Omega_i$ for $i=1,\,2$) since the
grand potential of a volume $V$ of mixed phase is found to be:
{\small
\begin{align}\label{mixedenergy}
\Omega_M=-P_MV \text{ with
}P_M=P_1\left[1-\frac{\left(\mu_2/\overline{\mu}_2-K\right)^2}{K^2-1}\right]
\end{align}
}
and the associated densities for the species are ($K\neq 1$):
\begin{subequations}\label{mixeddens}
\begin{align}
n_{1M}&\equiv
\overline{n}_1\left(\frac{1-K[\mu_2/\overline{\mu}_2]}{1-K^{2}}\right),\\
n_{2M}&\equiv\overline{n}_2\left(\frac{\mu_2/\overline{\mu}_2-K}{1-K^{2}}\right).
\end{align}
\end{subequations}
When $K$ exceeds the value $\mu_2/\overline{\mu}_2$, pure phase
$1$ is the ground state. Its grand potential is indicated by the horizontal thick line in
Fig.~\ref{fig0a}. It is clear from expression~\eqref{mixedenergy}
that the grand potentials of pure phase $1$ and the
mixed phase coincide when $K=\mu_2/\overline{\mu}_2$; however, at that very point $n_{2M}$ vanishes and therefore, there is no two-phase coexistence. Instead, the transition from pure phase $1$ to the mixed phase is {\em critical}, with continuous first derivative of the grand potential, as can be seen from the curved dotted line in Fig.~\ref{fig0a} which is the mathematical continuation of
the grand potential of the mixed phase. A mixed phase does not exist in the interval $\mu_2/\overline{\mu}_2 < K <\overline{\mu}_2/\mu_2$, which can be seen from inspection of the signs of the densities in~\eqref{mixeddens}. Furthermore, when
$K>\overline{\mu}_2/\mu_2$, a mixed state (line with closely spaced dots) in
Fig.~\ref{fig0a} can again be identified, but it is unstable, as is easily derived from a
stability analysis. Its grand potential is even higher than that of the metastable pure phase 2.

Consider now the case of bulk two-phase coexistence $\mu_2=\overline{\mu}_2$. Then, the two open dots indicated in Fig.~\ref{fig0a} merge at $K=1$ so that pure phases $1$ and $2$ coexist whenever $K\geq 1$~\cite{ao}. A remarkable transition occurs at $K=1$: First, when going from $\mu_2\neq\overline{\mu}_2$ to $\mu_2=\overline{\mu}_2$ the character of the demixing transition changes from critical to first-order. Secondly, an infinite degeneracy occurs due to a rotational
symmetry in the GP equations,
\begin{align}\label{rotatie}
\widetilde{n}_1+\widetilde{n}_2=1,
\end{align}
as is readily seen by taking $K=1$ in 
Eqs.~\eqref{GP2}~\footnote{Moreover, an SU(2) invariance arises when
working with the field operators~\cite{kuklov}.}. We stress this
degeneracy at $K=1$, because further on, when studying
wetting transitions, we find a similar degeneracy for
inhomogeneous systems at wetting and for $K > 1$. 

\begin{figure}
\begin{center}
	\includegraphics[width=0.45\textwidth]{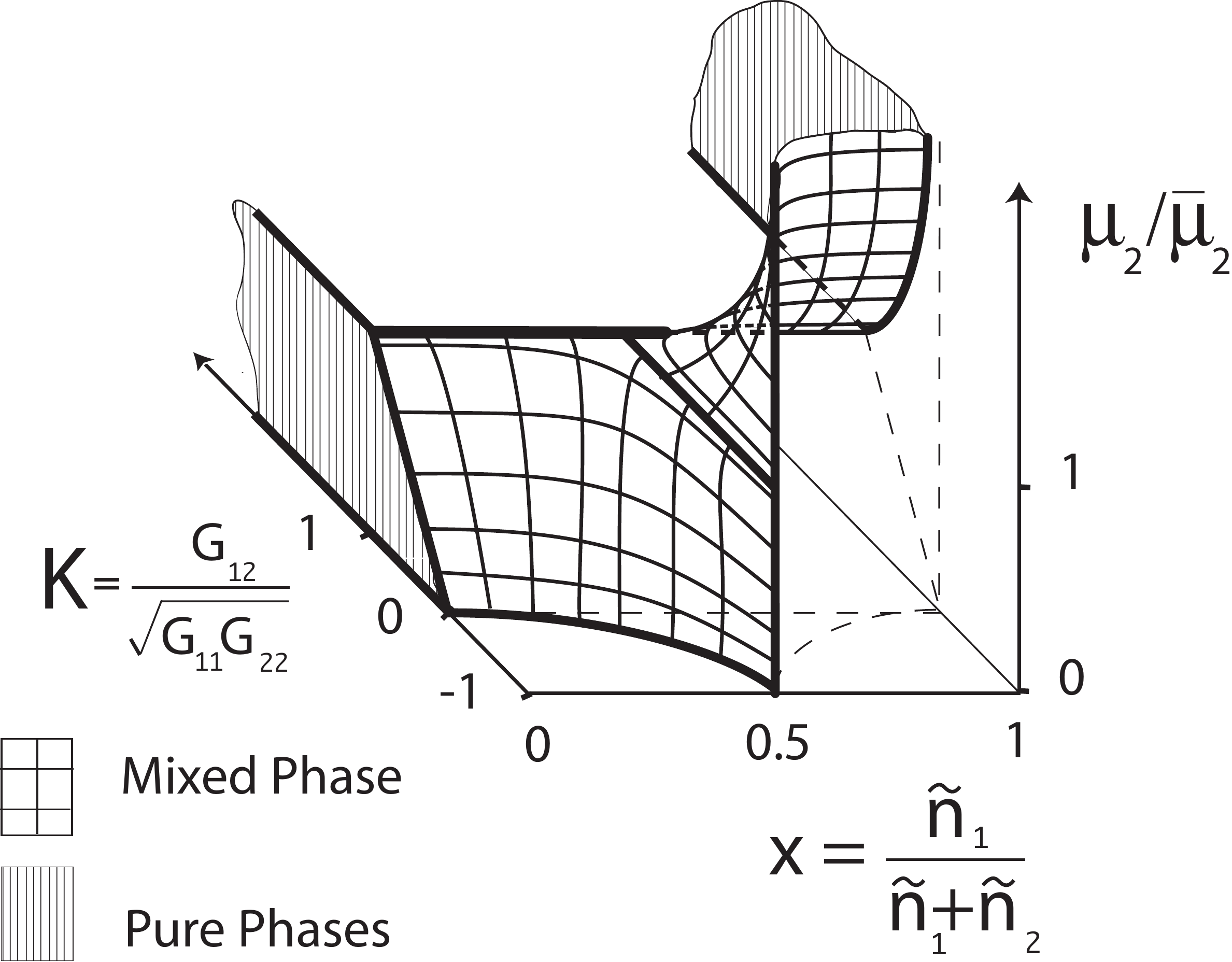}
       \caption{Bulk phase diagram as a function of
       $K$, $\mu_2/\overline{\mu}_2$ and the fraction
       $x=\widetilde{n}_1/(\widetilde{n}_1+\widetilde{n}_2)$. Hatched are the ground states. Pure phase $1$ ($2$) constitutes the ground state when
       $x=1$ ($x=0$) and $K>\mu_2/\overline{\mu}_2$
       ($K>\overline{\mu}_2/\mu_2$). Hence pure phases $1$ and $2$  coexist for $\mu_2/\overline{\mu}_2=1$ and $K\geq
       1$. Upon approach of the ``triple point" $K=1$ and $\mu_2/\overline{\mu}_2=1$, pure phases 1 and 2 and the mixed phase with $x=0.5$ coexist. Moreover, the triple point itself possesses an infinite (continuous) degeneracy in that $x$ can take all real values between zero and one.\label{fig0b}
       }
  \end{center}
\end{figure}

In Fig.~\ref{fig0b}, we give the bulk
$(x,\mu_2/\overline{\mu}_2,K)$ phase diagram, where
$x=\widetilde{n}_1/(\widetilde{n}_1+\widetilde{n}_2)$\label{defx}~\footnote{Note
that $x$ indicates the concentration at each point of the volume
and \textit{not} the volume fraction; we work at fixed chemical
potentials instead of fixed particle numbers.} denotes the
density fraction of species $1$. This phase diagram is analogous to the more
familiar $(x,P_2/P_1,T)$ phase diagrams for ordinary binary mixtures of fluids. The thick lines
and the hatched regions denote the ground states. Bulk coexistence
occurs between pure phases $1$ and $2$ when
$\mu_2=\overline{\mu}_2$ and $K > 1$ and three-phase coexistence of two pure phases ($x=0$ and $x=1$) and a mixed phase ($x=0.5$)
occurs when $\mu_2 =\overline{\mu}_2$ and $K\rightarrow 1$. We conclude
that wetting by a {\em pure phase} can be studied for $K>1$. Note that for
the mixed phase, stability is possible even for negative
$G_{12}$, down to $-\sqrt{G_{11}G_{22}}$, corresponding to $K = -1$.

\section{Surface Energy Excesses}\label{surt_sec_surtsurfacepotential}
For a semi-infinite system with translational symmetry in the $x$-$y$ plane, in which the atoms of species $i$ are bounded by the
trapping potential $U_i(\widetilde{z})$ centered about $\widetilde{z} = 0$, we now define the excess quantities. We assume that a steep ``wall" is present at $\widetilde{z}\approx 0$ which confines the particles mainly to $\widetilde{z} >0$. However, this need not be a hard wall and some particles may be found at $\widetilde{z}<0$. In the GCE there exists, up to a constant term, only one definition for the \textit{excess grand potential per unit area}, $\gamma$. Assuming the presence of pure phase $1$ in our
system for $\widetilde{z}\rightarrow \infty$, it is obtained by subtracting from the total grand potential
$\Omega$ the grand potential of a half space
($\widetilde{z}>0$) filled with pure phase $1$, both divided by
the area of the $x$-$y$ surface. With use of the GP
equations~\eqref{GP2} this yields:
\begin{align}\label{interfacetension}
\gamma_{_{W1}}=&\,\lim_{L\rightarrow \infty} \left [-2P_1\xi_1\int_{-\infty}^{ L}\text{d}\widetilde{z}\,\left(
\frac{\widetilde{\psi}^{4}_1}{2}+\frac{\widetilde{\psi}^{4}_2}{2}+K\widetilde{\psi}^{2}_1\widetilde{\psi}^{2}_{2}\right)\right.\nonumber\\
&\left.\quad\quad+P_1\xi_1\int_{0}^{L}\text{d}\widetilde{z}\right].
\end{align}
This expression allows us to define the \textit{surface tension}
or \textit{wall tension} $\gamma_{_{W1}}$ of condensate $1$
against a wall as the excess grand potential per unit
area~\eqref{interfacetension} for a semi-infinite system with
translational symmetry in the $x$-$y$ direction, for which the
following boundary conditions are satisfied:
\begin{subequations}
\begin{align}
&\widetilde{\psi}_1(\widetilde{z}\rightarrow -\infty)=0,\quad
\widetilde{\psi}_1(\widetilde{z}\rightarrow \infty)=1;\\
&\widetilde{\psi}_2( \widetilde{z}\rightarrow -\infty)=\widetilde{\psi}_2(\widetilde{z}\rightarrow \infty)=0.
\end{align}
\end{subequations}
Note that \eqref{interfacetension} only picks up (finite) surface and interface contributions, since the contributions from bulk-like regions cancel. Also note that, with this definition, $\gamma_{_{W1}}$
can be negative (e.g., for a steep wall positioned at some
$\widetilde{z} < 0$), however, without leading to any instability.
Likewise, we define the surface tension or wall tension of pure phase $i$ as the following excess energy per unit area,
\begin{align}\label{interfacetensionpure}
\gamma_{_{Wi,\mbox{pure}}}=\,\lim_{L\rightarrow \infty} \left[ -P_1\xi_1\int_{-\infty}^{L}\text{d}\widetilde{z}\,
\widetilde{\psi}^{4}_i
+P_i\xi_1\int_{0}^{L}\text{d}\widetilde{z}\right],
\end{align}
where the prefactors $P_1\xi_1$ and $P_i\xi_1$ are consistent with the scalings of $\psi$ and $z$ introduced in section II. For this excess energy the following boundary conditions and bulk condition are assumed (with $i \neq j$):
\begin{align}
\widetilde{\psi}_i(\widetilde{z}\rightarrow-\infty)=0,\quad
\widetilde{\psi}_i(\widetilde{z}\rightarrow\infty)=1;\quad
\widetilde{\psi}_j(\widetilde{z})=0.
\end{align}

At bulk two-phase coexistence of pure phases 1 and 2 ($P_1 = P_2$), we can define also the \textit{interfacial tension} $\gamma_{_{12}}$ as
the excess grand potential per unit area~\eqref{interfacetension}, but with the lower limit of the second integral in \eqref{interfacetension} extended to $-\infty$,
for an infinite system with translational symmetry in the $x$-$y$
direction, for which the boundary conditions are:
\begin{subequations}
\begin{align}
\widetilde{\psi}_{1}(\widetilde{z}\rightarrow-\infty)=\widetilde{\psi}_{2}(\widetilde{z}\rightarrow\infty)&=0,\label{interfbounda}\\
\widetilde{\psi}_{1}(\widetilde{z}\rightarrow\infty)=\widetilde{\psi}_{2}(\widetilde{z}\rightarrow-\infty)&=1.\label{interfboundb}
\end{align}
\end{subequations}

\section{Wetting at a Hard Wall}\label{sec_wettingsetup}
 We focus first on the standard wetting geometry for two BEC species first studied in \cite{indekeu}, i.e., the bosonic atoms are allowed to move freely in the
half space $\widetilde{z}>0$ but are blocked at
$\widetilde{z}=0$ by a hard wall. The hard wall gives rise to
Dirichlet conditions $\widetilde{\psi}_1(0)=0$ and
$\widetilde{\psi}_2(0)=0$. Also, infinitely far
from the wall, at $\widetilde{z}\rightarrow\infty$, we impose pure phase
$1$. For examining wetting, it suffices to consider densities that are inhomogeneous only in the direction perpendicular to the wall, so that $\widetilde{\psi}_1$ and $\widetilde{\psi}_2$
depend only on the coordinate $\widetilde{z}$. The boundary
conditions for partial wetting states are~\footnote{In the following, we tacitly assume that
gravity affects the condensate only on a much larger length scale than
the healing length. Indeed, by a rescaling of the Schr\"odinger
equation~$(\hslash^2/2m)d^2\psi/dz^2=-mgz\psi$ with $g$ the
gravitational constant, one arrives at the length of variation
$\lambda_{{gr}}=[\hslash^2/(2gm^2)]^{1/3}$. For rubidium
$\lambda_{{gr}}\approx 10^5$\AA $\,\gg\,\xi\approx
4000$\AA~\cite{leggett}.}:
\begin{align}\label{boundcond}
\widetilde{\psi}_1(0)=\widetilde{\psi}_2(0)=\widetilde{\psi}_2(\widetilde{z}\rightarrow\infty)=0\text{
and } \widetilde{\psi}_1(\widetilde{z}\rightarrow\infty)=1.
\end{align}
Note that for complete wetting states the boundary conditions are different in that pure phase 2 extends from $\widetilde{z}=0$ to $\infty$. Nevertheless, beyond this phase a 2-1 interface is ``inserted" so that the ultimate bulk phase is again pure phase 1, as in \eqref{boundcond}.

The essential quantities determining the wetting behavior are the surface tensions. One can easily obtain the hard wall tension
$\gamma_{_{Wi,\mbox{pure}}}$ and find that it is linear in $\xi_i$. Indeed, consider the half space $\widetilde{z}>0$ to be filled with condensate $i$. The GP Eqs.~\eqref{GP2a} and \eqref{GP2b}, together with a Dirichlet boundary condition at
$z=0$ (hard wall) yield, respectively, the profiles:
\begin{align}\label{surt_tanh}
\widetilde{\psi}_{1}=\tanh\left(\frac{z}{\sqrt{2}\xi_1}\right), \nonumber \\
\widetilde{\psi}_{2}=\sqrt{\frac{\mu_2}{\overline{\mu}_2}}\,\tanh\left(\frac{z}{\sqrt{2}\xi_2}\right).
\end{align}
From expression~\eqref{interfacetensionpure}, the associated hard wall
tension $\gamma_{_{Wi,\mbox{pure}}}$ is \cite{fetter}:
\begin{align}\label{surt_wallen}
\gamma_{_{Wi,\mbox{pure}}}=4\sqrt{2}P_i\xi_i/3,
\end{align}
which is mathematically similar to the tension of a normal-superconducting interface in the limit of strongly type I superconductors \cite{GL}.

At bulk two-phase coexistence ($P_1=P_2$), preferential adsorption of species $2$ therefore arises when
$\gamma_{_{W2}}<\gamma_{_{W1}}$ or, equivalently, when
$\xi_2<\xi_1$. For this reason, one can think of $\xi_1/\xi_2-1$ as
being the \textit{surface field} which, when positive, favors species $2$. Note that the healing lengths composing the surface field are themselves actually
\textit{bulk} parameters, as is clear from~\eqref{healing}.

\subsection{The expected behavior}\label{ssec_expect}
We assume the mixture is at bulk two-phase coexistence. The interfacial tension $\gamma_{_{12}}$ depends strongly on $K$ and
one may distinguish the following four regimes:\\
 {\bf A) } In the limit of \textit{strong segregation} or $1/K\rightarrow 0$, the two species
will have only a small spatial overlap so that
$\gamma_{_{12}}\approx \gamma_{_{W1}}+\gamma_{_{W2}}$. It can
readily be
checked that this inhibits complete wetting since under these circumstances the inequality $\gamma_{_{W1}} < \gamma_{_{W2}} + \gamma_{_{12}}$ (partial wetting) cannot become an equality.\\
 {\bf B) }  When again $1/K\rightarrow 0$ and in addition $\xi_2/\xi_1\rightarrow 0$ (strong surface field), in such a way
that $(\xi_2/\xi_1)\sqrt{K}$ remains finite, one finds that
$\gamma_{_{12}}=\gamma_{_{W1}}+\gamma_{_{W2}}-4P_1\xi_2\mathcal{G}([\xi_2/\xi_1]\sqrt{K})$
where the positive dimensionless function $\mathcal{G}$ typically takes values of order unity~\cite{BVS1}. The
condition for partial wetting becomes $\gamma_{_{W2}} >
2P_1\xi_2\mathcal{G}([\xi_2/\xi_1]\sqrt{K})$, from which one may
conclude that a transition from partial wetting to complete wetting is possible provided $ \mathcal{G}$ takes the value $2\sqrt{2}/3$ for some value of its argument. Thus we anticipate that the wetting phase boundary is parabolic for $1/K \rightarrow 0$, in the manner
\begin{align}\label{strongcond}
1/K\propto [\xi_2/\xi_1]^{2}.
\end{align}
 {\bf C) } Close to the demixing point where $K\approx 1$, the interface is
characterized by large interspecies penetration depths
$\Lambda_i\equiv\xi_i/\sqrt{K-1}$\label{lambdadeff}. It was found
in Refs.~\cite{mazets,malomed,barankov,ao} that, therefore, the
interfacial tension scales as $P\xi\sqrt{K-1}$. The vanishing
of $\gamma_{_{12}}$ when $K\rightarrow 1$ indicates that complete wetting is
unavoidable and that the wetting transition occurs,
according to Young's equation, when:
\begin{align}\label{weakcond}
K-1 \propto (1-\xi_2/\xi_1)^2,
\end{align}
with a proportionality constant of order unity,
implying a parabolic phase boundary about $K=1$ and $\xi_2/\xi_1 = 1$.
Actually, what happens near the (degenerate) triple point $K=1$ is
reminiscent of ``critical-point wetting'', with $1/K$ playing the
role of temperature, due
to the vanishing of $\gamma_{_{12}}$ for $K\rightarrow 1$.\\
 {\bf D) } In the case $\xi_2/\xi_1>1$, species $1$ is preferentially adsorbed at the wall. Then, assuming pure phase $2$ as the bulk
phase, the condition for \textit{complete drying} (CD) becomes
\begin{equation}\label{dewetting}
\gamma_{_{W2}} =\gamma_{_{W1}} + \gamma_{_{12}}.
\end{equation}
Fully analogously to cases {\bf A},
{\bf B} and {\bf C}, one finds that partial drying (PD)
is expected when $K\rightarrow \infty$, and complete drying (CD)
when $K\rightarrow 1$ and the transition from PD to CD occurs in
the region where~\eqref{strongcond} and~\eqref{weakcond} apply,
however, with $\xi_1$ and $\xi_2$ interchanged.
\begin{figure}
\begin{center}
	\includegraphics[width=0.45\textwidth]{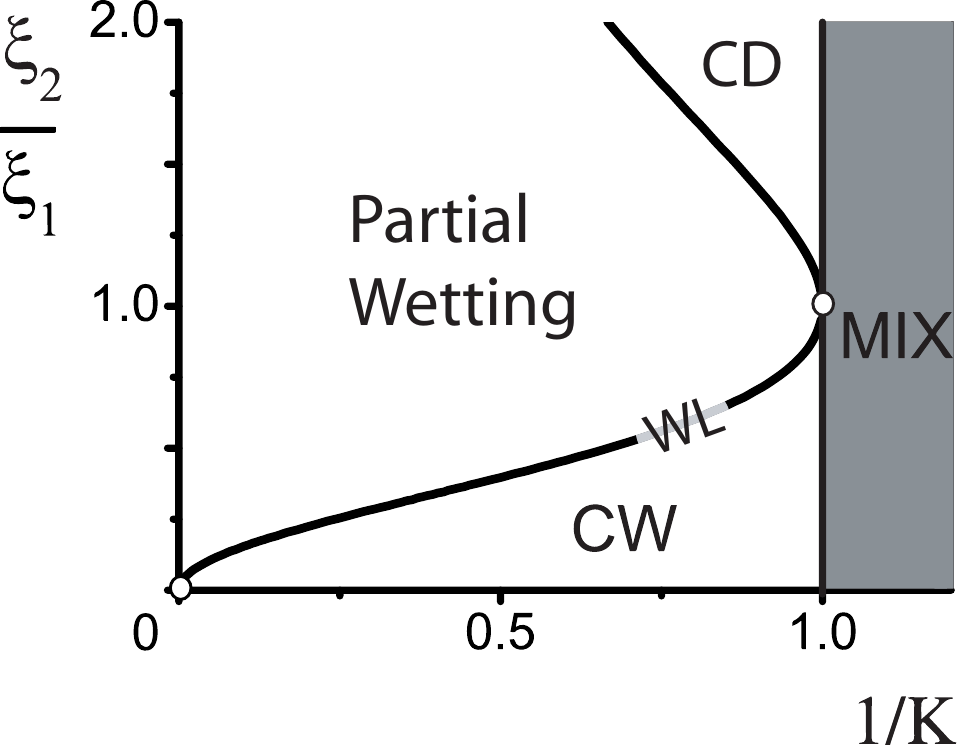}
       \caption{
       Wetting phase diagram in the plane of surface field $\xi_1/\xi_2 -1$ (represented here by the ratio  $\xi_2/\xi_1$) and
       reciprocal interaction strength $1/K$, 
       at bulk two-phase coexistence. A first-order
       phase boundary or wetting line (WL) separates partial wetting (PW) from complete wetting (CW), for $\xi_2/\xi_1 < 1$.
       For $\xi_2/\xi_1 > 1$ the roles of the condensates are interchanged and one may use
       ``drying'' in place of ``wetting''. For example, for $\xi_2/\xi_1 > 1$, partial wetting (see figure) signifies partial drying, and CW is replaced with CD. The phase boundary is
       parabolic in $1/K$ and $1/K -1$ as a function of $\xi_2/\xi_1$ and $\xi_2/\xi_1-1$, respectively, near the points $(0,0)$ and $(1,1)$.\label{fig1}
       }
  \end{center}
\end{figure}
\subsection{Phase diagram at bulk two-phase coexistence}\label{ssec_diagr}
In figure~\ref{fig1}, we show the exact wetting phase diagram at bulk two-phase 
coexistence as a function of $1/K$ and $\xi_2/\xi_1$. This figure
confirms all expectations expressed in {\bf A} through {\bf D} of the foregoing subsection. We now argue that at bulk coexistence, roughly speaking two
surface regimes are possible in equilibrium: Either an infinitely thick (macroscopic) layer of the wetting species 2 is
adsorbed or no atoms of that species are adsorbed. The parameter regimes in the
$(1/K,\xi_2/\xi_1)$-plane where PW and CW occur turn out to be
separated by a \textit{first-order wetting line} (WL), which at
each point has an infinite degeneracy \cite{indekeu,BVSPhD}. This degeneracy can to some extent be thought of as
a ``continuation to inhomogeneous states" of the degeneracy at the bulk triple point $K=1$.
We also show that WL (for $\xi_2/\xi_1 \leq 1$) is exactly given by the analytical expression (see also the second paper of \cite{BVS1}):
\begin{align}\label{nucleatiecond}
\sqrt{K-1}=\frac{\sqrt{2}}{3}\left[\frac{1}{\xi_2/\xi_1}-\xi_2/\xi_1\right].
\end{align}

First of all, as a function of the wall
tensions~\eqref{surt_wallen} and the wave functions of a $1$-$2$
interface, $\widetilde{\psi}_1$ and $\widetilde{\psi}_2$, which obey
boundary conditions~\eqref{interfbounda} and~\eqref{interfboundb},
the condition for partial wetting can be rewritten with use of
Eq.~\eqref{behoudenergie}:
\begin{align}\label{voorwaarde}
\int_{-\infty}^{\infty}{\text{d}\widetilde{z}\,\left(
\dot{\widetilde{\psi}^{2}_1}+[\xi_2/\xi_1]^2\dot{\widetilde{\psi}^{2}_2}\right)} > \frac{\sqrt{2}}{3}(1-\xi_2/\xi_1).
\end{align}
Remarkably, all constituents of this condition, even the profiles
$\widetilde{\psi}_1$ and $\widetilde{\psi}_2$, are fully
determined by $K$ and $\xi_2/\xi_1$ (see GP Eqs.~\eqref{GP2}). By
numerical integration, we have verified that the inequality~\eqref{voorwaarde} becomes an equality (i.e., Antonov's rule, valid for complete wetting)
for values of $K$ and $\xi_2/\xi_1$ which satisfy the exact
relation~\eqref{nucleatiecond}.

What kind of behavior can one expect
close to this wetting transition line? One possibility is that films of finite thickness appear as premonitory surface states initiating the interface delocalization or ``wetting" transition. Such films may be nucleated as infinitesimal films through a critical nucleation transition. In order to study this possibility we linearize
Eq.~\eqref{GP2b} in terms of the wave function
$\widetilde{\psi}_2$, about $\widetilde{\psi}_2 =0$. This allows one to study exactly the infinitesimal nucleation of
species $2$ when species $1$ occupies the entire half space:
\begin{figure}
\begin{center}
	\includegraphics[width=0.22\textwidth]{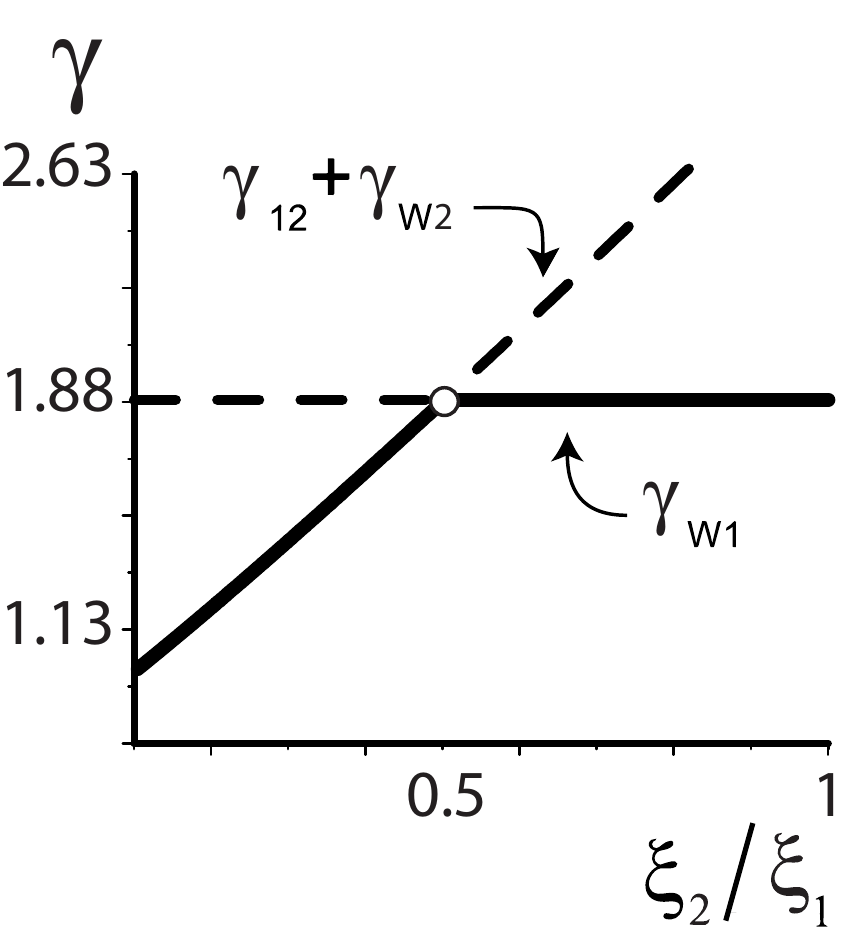}
	\includegraphics[width=0.22\textwidth]{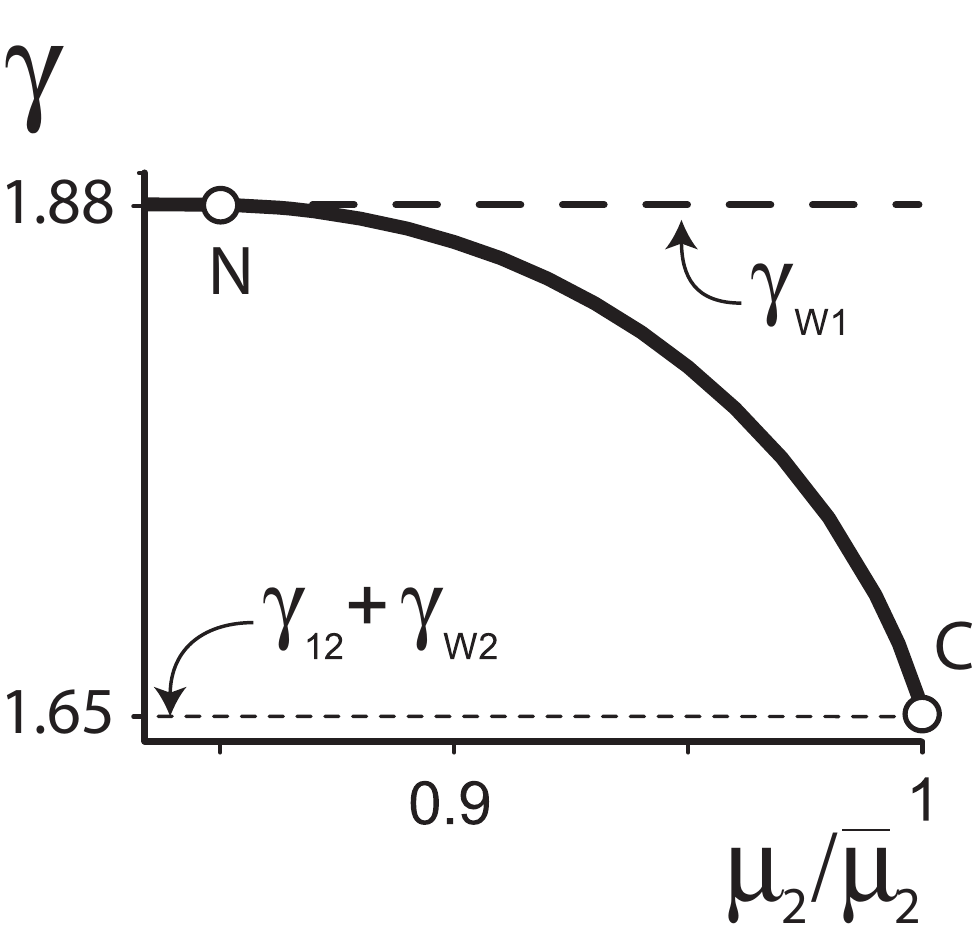}
        \caption{
        (Left) The excess grand potential per unit area $\gamma$ in units of $P_1\xi_1$ against the ratio of lengths inherent to the
        surface field, $\xi_2/\xi_1$, for fixed $K=1.5$, at bulk two-phase coexistence. We vary $\xi_2$
        and keep $\xi_1$ fixed.
        The ground state energy is indicated by the thick line. Clearly, the wetting transition at $\xi_2/\xi_1=0.5$ is of first
        order: The excess energy of pure phase $1$ adsorbed at the wall
        cuts the excess energy of a system in which an infinite layer of species $2$ wets the
        wall. (Right) The excess grand potential per unit area $\gamma$ in units of $P_1\xi_1$ of a
       prewetting film, which is nucleated with infinitesimal amplitude at N and becomes macroscopically thick at C.
       The energy is shown for $\overline{\xi}_2/\xi_1=2/5$ and $K=1.5$ as a function of the field variable which can be used to measure the deviation from bulk two-phase coexistence, $\mu_2/\overline{\mu}_2$. The thick line denotes the ground state energy. The prewetting transition is critical: At the nucleation point (N)
       an infinitesimal film of species $2$ is nucleated at the
       wall. 
        \label{fig2}
        }
  \end{center}
\end{figure}
\begin{align}\label{nlvgl1}
[\xi_2/\xi_1]^2\ddot{\widetilde{\psi}}_2=-\widetilde{\psi}_2+K\tanh^{2}(\widetilde{z}/\sqrt
{2})\widetilde{\psi}_2.
\end{align}
The solutions for $\widetilde{\psi}_2$ must fulfill the boundary
conditions~\eqref{boundcond} and we find that they decay exponentially for large $z$. They correspond to films of
thickness (decay length) equal to half the previously introduced penetration depth, $\Lambda_2/2=\xi_2/(2\sqrt{K-1})$. (The factor 1/2 comes from squaring the wave function for obtaining the density.) The wave
functions are exactly given by (see Appendix A):
\begin{align}\label{oplossing}
\widetilde{\psi}_2=\epsilon \tanh(\widetilde{z}/\sqrt
{2})[\cosh(\widetilde{z}/\sqrt
{2})]^{-\sqrt{2}\xi_1/\Lambda_2},
\end{align}
where $\epsilon$ is by assumption an infinitesimal amplitude.
As explained in Appendix A, nucleation of this kind
can only exist when condition~\eqref{nucleatiecond} is satisfied. Consequently, in the wetting phase diagram, at bulk two-phase coexistence, the nucleation line for infinitesimal adsorbed films {\em coincides} with the wetting phase boundary WL. This is extraordinary. Moreover, at each
point on WL, not only do nucleated infinitesimal films and layers of infinite
thickness solve the GP equations, but also layers of {\em arbitrary finite thicknesses} exist as solutions, all at the same value of the grand potential. 

This degeneracy of the grand potential corresponds to a one-parameter family of minima that form 
a ``gutter" in the $(\widetilde{\psi}_1(x),\widetilde{\psi}_2(x))$
function space, when the grand potential is plotted as a function of that parameter (i.e., the adsorption defined later in \eqref{adsorption}) and a second, independent parameter. Note that the decay length of nucleated infinitesimal films, $\Lambda_2/2$,
diverges when $K \rightarrow 1$ since it is proportional to $1/\sqrt{K-1}$.
Therefore, there appears to be a connection between the degeneracy of bulk phase densities encountered at the bulk triple point (see \eqref{rotatie}) and the degeneracy found here for the limiting densities of inhomogeneous states for $\widetilde{z} \rightarrow \infty$.

The crossing of the excess energies of PW and CW states, conspicuous in
Fig.~\ref{fig2} (Left), proves that the wetting phase boundary WL in Fig.~\ref{fig1} is a
first-order line. One can ask which obvious physical quantity displays a jump across this line. To answer this we imagine traversing the wetting line in Fig.~\ref{fig1} at constant surface field $\xi_2/\xi_1$ and varying $K$. We consider the first derivative of the excess grand potential with respect to $K$ and readily observe, using e.g.~\eqref{interfacetension}, that this quantity corresponds to the {\em overlap} of the condensate densities of the two species,
\begin{equation}\label{overlapping}
\frac{\partial \gamma}{\partial K} = -2P_1\xi_1\int_{-\infty}^{\infty}\text{d}\widetilde{z}\,
\widetilde{\psi}^{2}_1\widetilde{\psi}^{2}_{2}.
\end{equation}
Upon approach of the wetting phase boundary from the PW regime the overlap is zero (because species 2 is absent), whereas the overlap is finite (assuming $K < \infty$) when the same point on the wetting phase boundary is approached from the CW regime. To calculate the overlap in the CW regime it suffices to consider the interface between phases 1 and 2. We conclude that the density overlap is a good order parameter for elucidating the first-order nature of this wetting transition.

\begin{figure}
\begin{center}
	\includegraphics[width=0.45\textwidth]{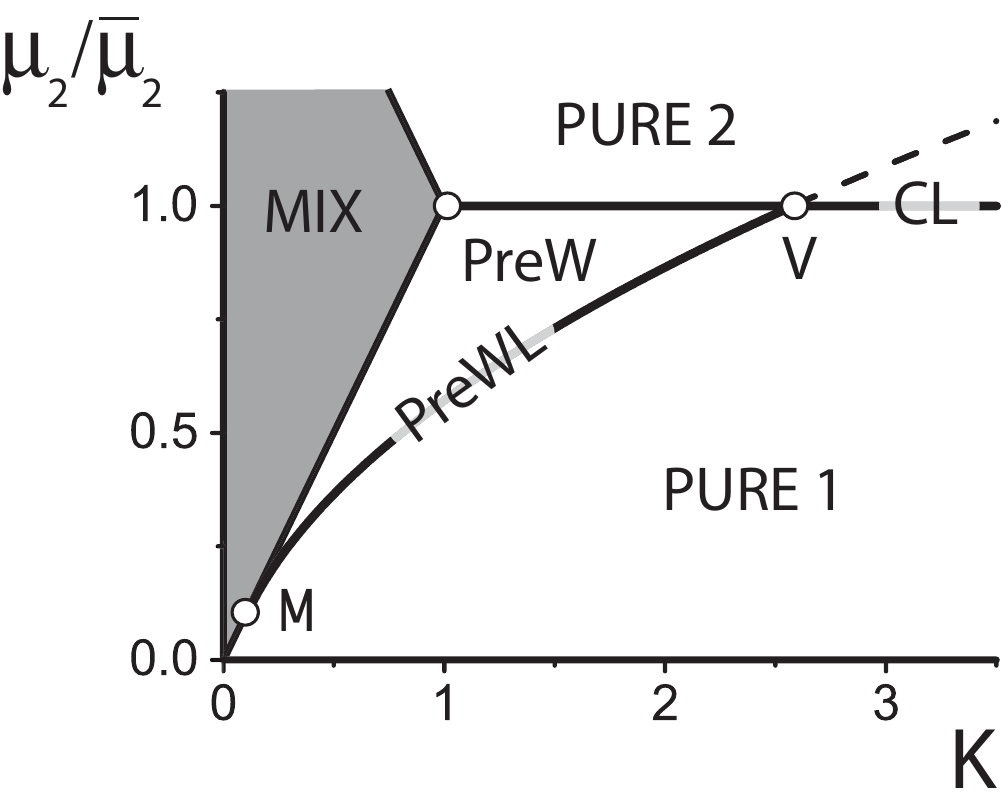}
        \caption{
        The prewetting phase diagram for $\overline{\xi}_2/\xi_1=1/3$ as a function of the deviation from
        bulk coexistence $\mu_2/\overline{\mu}_2-1$ and the relative interaction parameter
        $K=G_{12}/\sqrt{G_{11}G_{22}}$. Prewetting (PreW) is found above the prewetting line
        (PreWL) and below the bulk coexistence line (CL).  Physically, at fixed $K>1$, moving vertically from a point on 
        PreWL to a point on CL means going from a nucleated to an infinitely
        thick adsorbed
        film. For $K<1$, upon entering the region of bulk mixed phase (MIX) from the PreW region,
        the growth of the wetting layer of species $2$ is preempted by the
        bulk nucleation of species $2$. That is, bulk phase $1$ becomes unstable before wetting is achieved. See Fig.~\ref{fig21} for more detail and various scenarios. \label{fig3}
        }
  \end{center}
\end{figure}
\subsection{Phase diagram off of coexistence}\label{sec_off}
As one decreases the chemical potential $\mu_2$ from its
value at two-phase coexistence $\overline{\mu}_2$, pure phase $2$
is no longer a bulk ground state ($\Omega_{{2}}=-P_2V$ increases
due to the minus sign). Nevertheless, while
pure phase $1$ is stable in bulk, the surface can still be {\em prewetted} by films of
finite thickness of species $2$. In what follows, we show that,
contrary to common expectations, the \textit{first-order} wetting transition at
coexistence has an extension in the form of a \textit{critical} transition off
of coexistence; moreover, the resulting prewetting line coincides
with the nucleation line for infinitesimal films of species 2.

\begin{figure*}
  \begin{center}
\includegraphics[width=\textwidth]{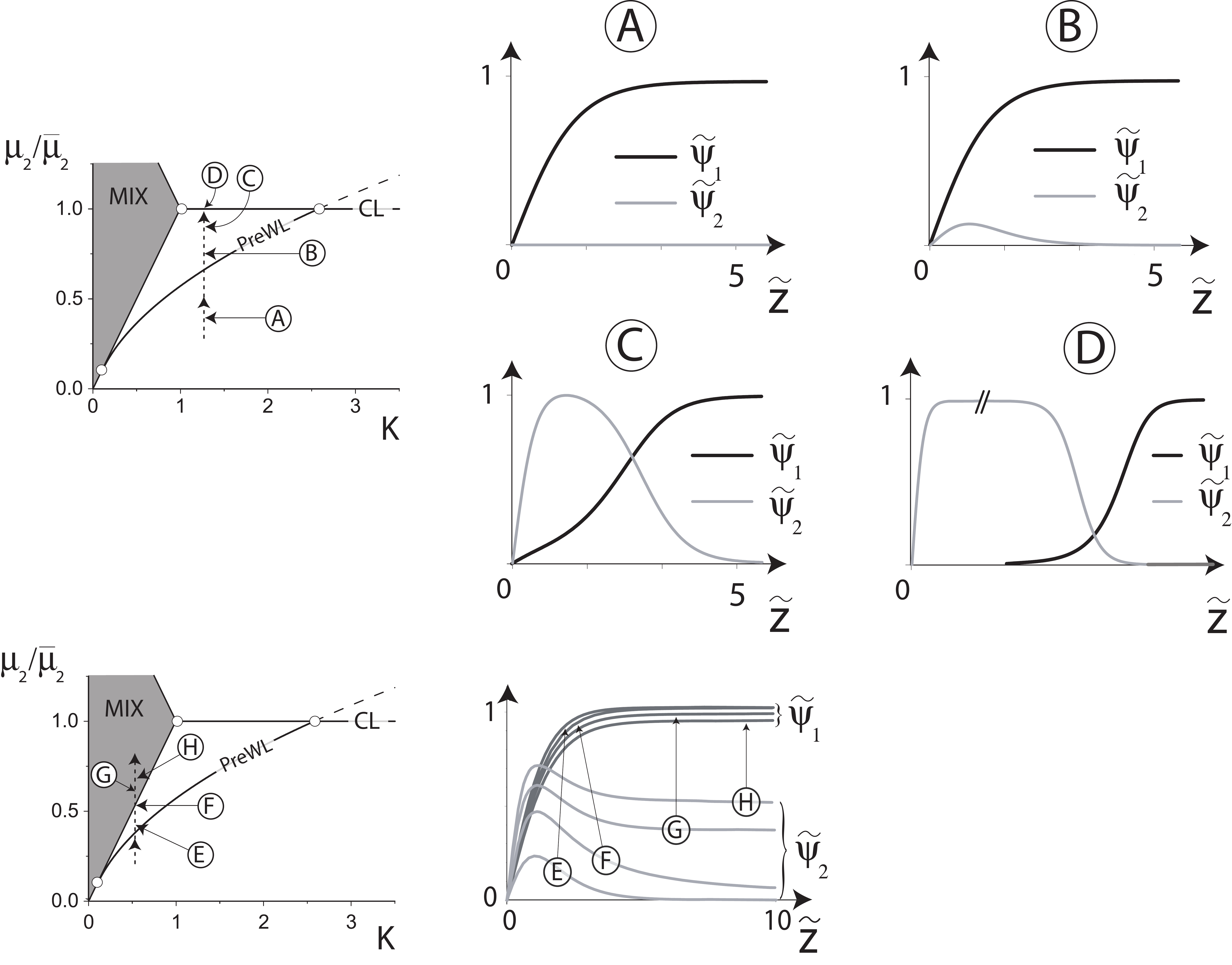}
        \caption{Prewetting states for $\overline{\xi}_2/\xi_1 = 1/3$. Panels A-D display profiles of the condensate wave
functions for the approach to complete wetting along the trajectory at $K>1$ (vertical dashed line)
indicated in the top left prewetting diagram. Note that panel D corresponds
to a complete wetting state in which an infinitely thick layer of phase $2$
is present between the hard wall and bulk phase $1$.  Profiles E-H depict the
wave functions of condensates $2$ (marked by the circles with letters E-H) and $1$ (indicated by the arrows emanating from the circles) along the trajectory at $K<1$ (vertical dashed line) indicated in the
bottom left prewetting diagram.   State E corresponds to a prewetting film; this film continuously
grows up to the point where  the bulk transition to the mixed phase is reached   
(state  F). Note that this film remains microscopically thin and does not become a wetting layer. From the profiles in state G one sees that there appears a
nonzero   bulk density of species $2$. Finally, for higher values of
$\mu_{2}/\overline{\mu}_{2}$, the bulk density of species $2$
increases (state H). 
        \label{fig21}}
         \end{center}
\end{figure*}
We start from the examination of nucleation
for a system at given $\mu_2/\overline{\mu}_2$, $\overline{\xi}_2/\xi_1$ and
$K$. Extending Eq.~\eqref{nlvgl1} to general values for the
chemical potentials, one obtains:
\begin{align}\label{nlvgl2}
[\overline{\xi}_2/\xi_1]^2\ddot{\widetilde{\psi}}_2=-[\mu_2/\overline{\mu}_2]\widetilde{\psi}_2+K\tanh^{2}
(\widetilde{z}/\sqrt{2})\widetilde{\psi}_2.
\end{align}
This can be readily transformed into the form of Eq.~\eqref{nlvgl1} and one concludes that, off of bulk coexistence,
infinitesimal nucleation occurs when $K$, $\overline{\xi}_2/\xi_1$ and
$\mu_2/\overline{\mu}_2$ are confined to the surface:
\begin{align}\label{PWnuclcond}
\sqrt{K-\mu_2/\overline{\mu}_2}=\frac{\sqrt{2}}{3}
\left[\frac{\mu_2/\overline{\mu}_2}{\overline{\xi}_2/\xi_1}-\overline{\xi}_2/\xi_1\right],
\end{align}
which of course reduces to Eq.~\eqref{nucleatiecond} for
$\mu_2=\overline{\mu}_2$. Nucleated films have a typical
thickness (decay length) of $\overline{\xi}_2/(2\sqrt{K-\mu_2/\overline{\mu}_2})$.

The nucleation constraint~\eqref{PWnuclcond} provides all necessary
information for drawing the prewetting phase diagram. A representative section is presented in Fig.~\ref{fig3}, calculated for fixed $\overline{\xi}_2/\xi_1=1/3$. The
prewetting line (PreWL) indicates the onset of nucleated infinitesimal films, as
given by expression ~\eqref{PWnuclcond}. The transition is critical.
In the prewetting region (PreW in Fig.~\ref{fig3}), thin
films grow thicker upon approach of the bulk coexistence line (CL). Importantly,
we find numerically that prewetting films are energetically favorable as
compared to states with no prewetting film. This is exemplified
by the case $\overline{\xi}_2/\xi_1=2/5$ and $K=1.5$, for which we show the
excess grand potential per unit area $\gamma$ in Fig.~\ref{fig2}
(Right). Clearly, $\gamma$ drops below the energy $\gamma_{_{W1}}$ (no film of phase 2)
at the nucleation point (N) and continues to decrease until the
point of bulk coexistence (C) is reached (macroscopic wetting layer of phase 2). Moreover, one notices that the departure of the
excess energy $\gamma$ away from the value $\gamma_{_{W1}}$
takes place via a critical transition.

Inspection of~\eqref{PWnuclcond} shows that the prewetting line PreWL starts
at the point M (see Fig.~\ref{fig3}) where
$\mu_2/\overline{\mu}_2=[\overline{\xi}_2/\xi_1]^2$ and where PreWL
tangentially meets the second-order bulk demixing line
$\mu_2/\overline{\mu}_2=K$ (cf. Fig. \ref{fig0a}). In point M of Fig.~\ref{fig3}
there is (critical) nucleation of phase $2$ in bulk. At the other end, in contrast with what is
commonly expected~\cite{hauge}, but nevertheless in full accord with surface thermodynamics, the line PreWL
cuts the coexistence line (CL) in point V under a non-zero angle \cite{indekeu}.

For low values of the surface
field, that is, when $\overline{\xi}_2/\xi_{1}\uparrow 1$, the prewetting
line of Fig.~\ref{fig3} shrinks and shifts upwards towards the bulk triple
point at $K=1$. For high values of the surface field (for $\overline{\xi}_2/\xi_1 \downarrow 0$), the prewetting region of Fig.~\ref{fig3}
grows as the points V and M move apart and away from the bulk triple point. 

One can understand the anomalous first-order character of the wetting transition, featuring an
infinite degeneracy at bulk coexistence, by taking a closer look at the
second-order prewetting transition. Indeed, as is seen from
Fig.~\ref{fig3}, the range over which $\mu_2/\overline{\mu}_2$
varies between onset of nucleation and divergence of the
prewetting layer on a trajectory of constant $K$, vanishes upon
approach of the point $V$. Since prewetting states are
energetically favorable, and all thicknesses must be realized in a prewetting segment of vanishing length (in the variable $\mu_2/\overline{\mu}_2$), a continuous degeneracy of film thicknesses must follow in the point $V$.

\begin{figure}
\begin{center}
			\includegraphics[width=0.45\textwidth]{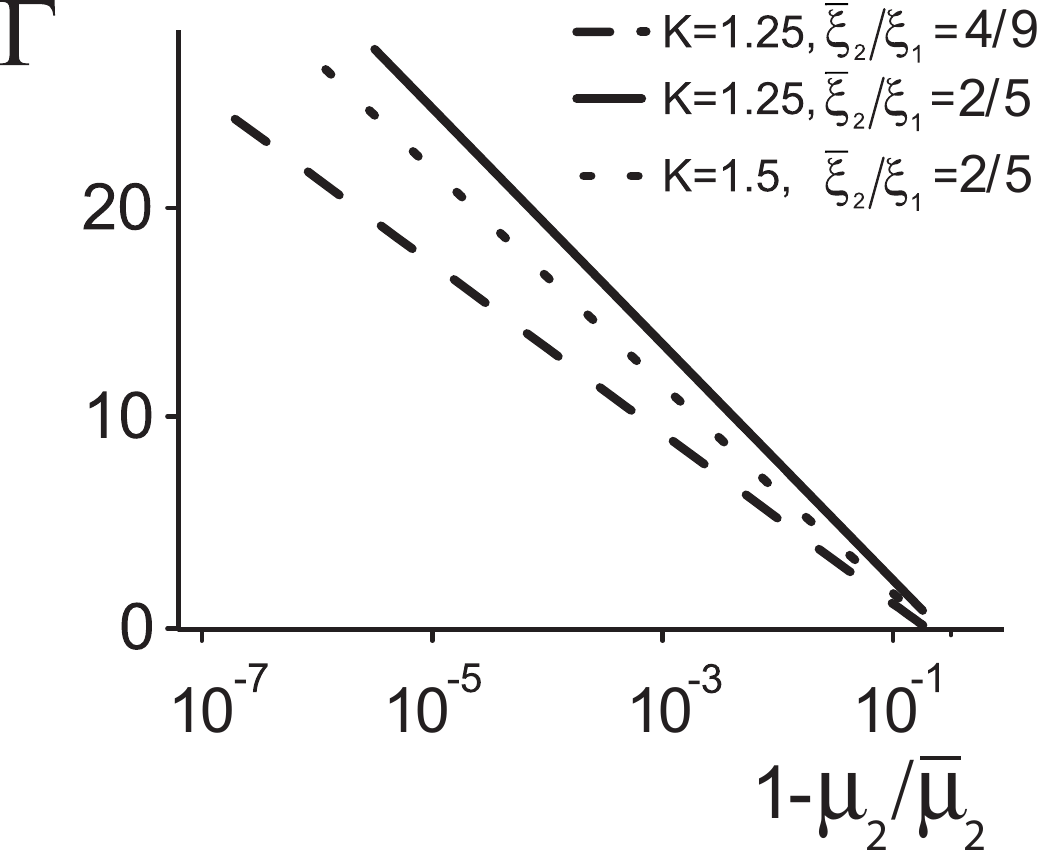}

       \caption{The numerically obtained logarithmic
       increase of the adsorption of species 2 (see~\eqref{adsorption}),
       associated with the prewetting film,
       as a function of $1-\mu_2/\overline{\mu}_2 = 1-\sqrt{P_2/P_1}$, which is proportional to the pressure difference $P_1-P_2$ when we are close to two-phase coexistence ($P_1=P_2$). The approach to complete wetting at bulk coexistence is accomplished for three different pairs of values of $K$ and $\xi_{2}/\xi_{1}$, in the prewetting regime.
       \label{fig9b}
        }
\end{center}
\end{figure}

We depict in Fig.~\ref{fig21} the density
profiles which are observed upon approach of the coexistence line
along a prewetting path at $K>1$ (panels A-D) and a path at $K<1$
(profiles E-H). A suitable measure of the thickness of the wetting layer is obtained through the (dimensionless) adsorption, which is proportional to the derivative of the surface excess energy with respect to the chemical potential. The adsorption of species 2 is defined as~\footnote{Here we ignore the distinction between $\mu_2$ and $\overline{\mu}_2$, since we are interested in the behaviour of the adsorption (very) close to bulk two-phase coexistence.}
\begin{align}\label{adsorption}
\Gamma=\int_{0}^{\infty}\text{d}\widetilde{z}\,\widetilde{\psi}_2^2.
\end{align}
For large values of $\Gamma$, the wetting layer thickness is proportional to $\Gamma$. The film thickness diverges
logarithmically upon approach of bulk two-phase coexistence, as we show in Fig.~\ref{fig9b}. This slow
divergence is expected for the approach to complete wetting in
systems with short-range interactions \cite{BonnRMP}.

\begin{figure}
\begin{center}
	\includegraphics[width=0.45\textwidth]{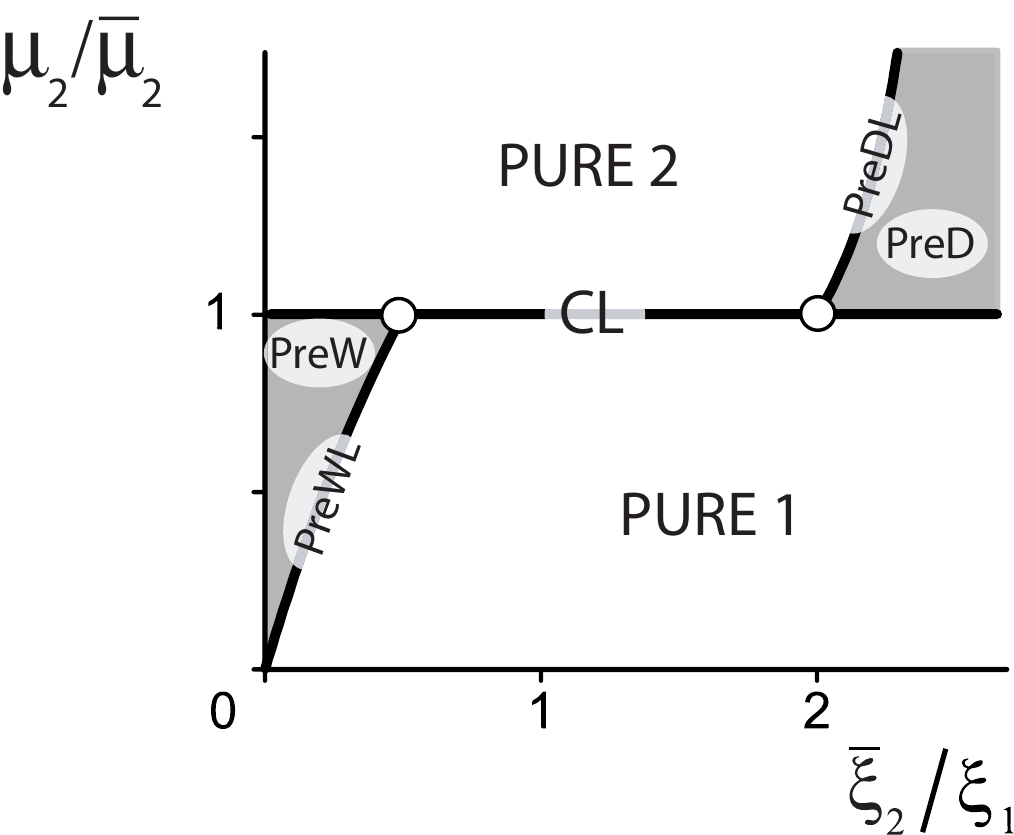}
       \caption{The prewetting phase diagram as a function of $\overline{\xi}_2/\xi_1$ and the deviation from bulk two-phase
       coexistence $\mu_2/\overline{\mu}_2-1$ for $K=1.5$. Prewetting (PreW)
       occurs in the shaded region on the left whereas ``predrying'' (PreD)
       (adsorption of species $1$ at the wall with pure phase $2$ in bulk)
       occurs in the shaded region on the right. Upon crossing the
       prewetting line (PreWL), or the predrying line (PreDL), a
       second-order surface transition occurs.
       \label{fig9a}
        }
\end{center}
\end{figure}

Finally, in
Fig.~\ref{fig9a}, the prewetting phase diagram is
drawn as a function of $\overline{\xi}_2/\xi_{1}$ and
the deviation from two-phase coexistence $\mu_{_2}/\overline{\mu}_{2}-1$
for $K=1.5$. For  
$\xi_{2}/\xi_{1} <1$ (positive surface field), phase $2$ is favored near the wall and the
prewetting line (PreWL) which connects the points
$(\overline{\xi}_2/\xi_{1},\,\mu_{_2}/\overline{\mu}_{2})= (0.5,\,1)$
and $(0,\,0)$, bounds a region where prewetting (PreW) by phase 2 occurs when
phase $1$ constitutes the bulk phase. For $\xi_{2}/\xi_{1} >1$ (negative surface field), the situation can be seen to be identical after
interchanging the two species: At the point $(2,\,1)$, being
the reciprocal of $(0.5,\,1)$, a predrying line (PreDL) starts
which bounds a predrying (PreD) region. In this region a film of
finite thickness of species $1$ is adsorbed at the wall while
phase $2$ constitutes the bulk phase.

\section{Wetting at a Soft Wall}\label{sec_soft}
In this section we assume that the BEC mixture is {\em at bulk two-phase coexistence}. In order to study wetting for an
experimentally relevant setup, we relax the hard wall. We do this
by taking the confining potential of species $i$ to be a ``soft
wall'', i.e.,~an exponentially decaying potential along the
$z$-direction:
\begin{align}\label{surface potential2}
U_i(z)=U_{i0}e^{-z/\lambda_i},
\end{align}
where $U_{i0}>0$. One recovers a hard wall for
$\lambda_i/\xi_i \rightarrow 0$. 

\subsection{Surface excess energies at a soft wall}\label{sec_onecondensate}
Before capturing the essence of a two-species semi-infinite system, confined on
one side by the soft potentials~\eqref{surface potential2}, we
first consider only species $i$ near the softened walls. We argue
that for small $\lambda_i/\xi_i$ it is justified to model the soft
wall by a {\em shifted hard wall}.

In the hard wall limit $\lambda_i/\xi_i\rightarrow 0$, the wave
function $\widetilde{\psi}_i$ has a $\tanh$ profile; relaxation of
$\lambda_i/\xi_i$ (away from zero) therefore affects the wave function only in the
vicinity of $\widetilde{z}=0$. One can prove that a surface
potential with a small onset ratio $\lambda_i/\xi_i$, gives rise
to a surface tension, derived in Appendix \ref{sec_softwall},
\begin{align}\label{corrsurface}
\gamma_{_{Wi, \mbox{pure}}}=\gamma_{_{i0}}+\gamma_{_{i1}}
+\mathcal{O}([\lambda_i/\xi_i]^5),
\end{align}
where
\begin{align}\label{softsurfacepotentialexpansions}
\begin{cases}
\gamma_{_{i0}}=4\sqrt{2} P_i\xi_i/3,\\
\gamma_{_{i1}}= P_{_i}\Delta_i,
\end{cases}
\end{align}
and we define:
\begin{align}\label{lengths}
\Delta_i
\equiv\lambda_{i}\left(\ln\left(
[U_{{i0}}/\mu_{_i}][\lambda_{i}/\xi_{i}]^2\right)+1.154\right. 
\left. -3.205[\lambda_{i}/\xi_{i}]^2\right),
\end{align}
\begin{figure}
\begin{center}
	\includegraphics[width=0.45\textwidth]{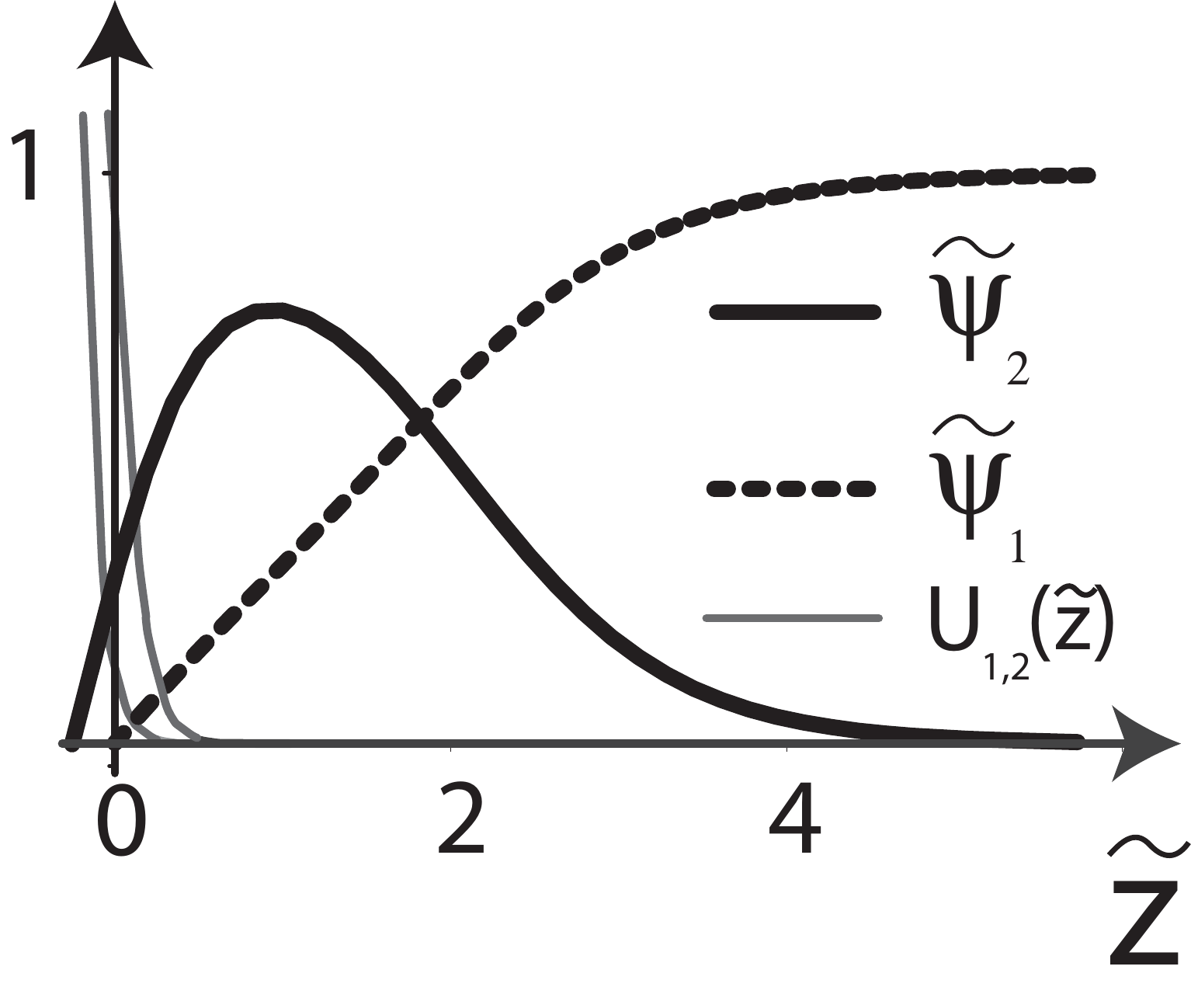}
      \caption{
      Illustration of the two condensate wave functions confined on one side by a soft wall. The exponential surface potentials are
      indicated by the thin gray lines. For the configuration above,
      one can see that the relative trap displacement satisfies $\Delta<0$, since phase $2$
      is shifted to the left. As a consequence, phase $2$ is more favored by the
      surface than it already was for the (unshifted) hard wall configuration, given that $\xi_{2} < \xi_{1}$.
       \label{fig15}
        }
\end{center}
\end{figure}
Note that $\gamma_{_{i1}}$ and all higher-order contributions vanish for the case of a hard wall boundary. This is consistent with the fact that $\gamma_{_{i0}}$ coincides with the hard wall surface tension given in Eq.~\eqref{surt_wallen}. The calculation reveals that the three terms in \eqref{lengths} arise \textit{because
of the shift of the $\tanh$-profile}, while the fifth-order term in~\eqref{corrsurface} also expresses the corrections due to wave-function distortions away
from the $\tanh$ profile. Provided that the fifth-order term can be neglected, the wave function
$\widetilde{\psi}_i$ again acquires the form of a $\tanh$-profile,
however, shifted away from the origin $\widetilde{z}=0$ over a
length $\Delta_i$. Consequently, one can model the soft wall by a
shifted hard wall. In case both particle species are present near
the soft walls, one can prove that it is again justified to
replace the soft walls with hard walls which are shifted over lengths
$\Delta_i$ as defined in expression~\eqref{lengths}. Therefore, the
shifts are not affected by the presence of an additional species. Fig.~\ref{fig15} illustrates the notion of a shifted hard wall boundary for the wave functions of two condensates adsorbed at soft walls.

\subsection{Phase Diagram For Soft Walls}\label{sec_phase diagrams}
In the previous subsection, we found out that shifted hard walls can replace the softer walls~\eqref{surface potential2} whenever $\lambda_i/\xi_i\ll 1$. By taking the length $\lambda_i \,(>0)$ small compared to the healing length $\xi_i$, we argue further that
only one parameter (instead of the initial four: $U_{10}$, $U_{20}$, $\lambda_1$ and $\lambda_2$) is
sufficient to characterize the system. Moreover, the sign of this
parameter plays an important role in determining whether the wetting transition is of first order or critical.
Taking the
new origin $z'=0$ at the position of the shifted hard wall of species
$1$ ($z=\Delta_1$), the \textit{relative trap displacement}
\begin{align}\label{gap}
\Delta\equiv\Delta_2-\Delta_1
\end{align}
expresses the new position $z' = \Delta$ of the shifted hard wall for species $2$
($z = \Delta_2$). 

One may now ask what are the
modifications to the wetting phase diagram at two-phase coexistence (Fig.\ref{fig1}) after
softening the wall. Since the $1$-$2$ interfacial tension is
independent of the confining surface potential, the condition for
complete wetting is easily found from Eqs.~\eqref{voorwaarde},
~\eqref{corrsurface},~\eqref{softsurfacepotentialexpansions} and~\eqref{gap} to be
($\widetilde{\psi}_1$ and $\widetilde{\psi}_2$ are the wave
functions of the $1$-$2$ interface):
 {\small
\begin{align}\label{CWcondition2}
\gamma_{12}=\int_{-\infty}^{\infty}{\text{d}\widetilde{z}\,\left(
\dot{\widetilde{\psi}^{2}_1}+[\xi_2/\xi_1]^2\dot{\widetilde{\psi}^{2}_2}\right)}
= \frac{\sqrt{2}}{3}(1-\xi_2/\xi_1)-\frac{\Delta}{4\xi_{1}}.
\end{align}
}

Consider the case $\Delta >0$. The nucleation of species $2$ is
determined by equation~\eqref{nlvgl1}, with $\widetilde{z} \rightarrow \widetilde{z}'$, together with the
boundary condition:
\begin{align}\label{bcn}
\widetilde{\psi}_2(z'=\Delta)=\widetilde{\psi}_2(z'\rightarrow\infty)=0,
\end{align}
where the solution $\widetilde{\psi}_2$ is given in
Eq.~\eqref{nucleationsolution} of Appendix A. On the other hand, for $\Delta<0$, nucleation of
adsorbed species $2$ is solved for by matching
the extrapolation lengths of the nucleated density profiles of
species $2$ at $z'=0$; we know that when $z'<0$,
$\widetilde{\psi}_2\propto \sin[(z'-\Delta)/\xi_2]$, while for
$z'>0$, the solution is given in
expression~\eqref{nucleationsolution}. Using
expression~\eqref{nucleationzero}, one straightforwardly
calculates that at nucleation:
\begin{align}\label{gammadef}
-\frac{\sqrt{2}\Gamma[(A^{-}+1)/2]\,\Gamma[(1+A^{+})/2]}{\Gamma[A^{+}/2]\,\Gamma[A^{-}/2]}=
\frac{\cot}{[\xi_2/\xi_1]}\left(\frac{\Delta}{\xi_2}\right)
\end{align}
where $A^{\pm}$ is defined in expression~\eqref{constanten} and
$\Gamma$ is the gamma function (not to be confused with the adsorption).

\begin{figure}
\begin{center}
	\includegraphics[width=0.45\textwidth]{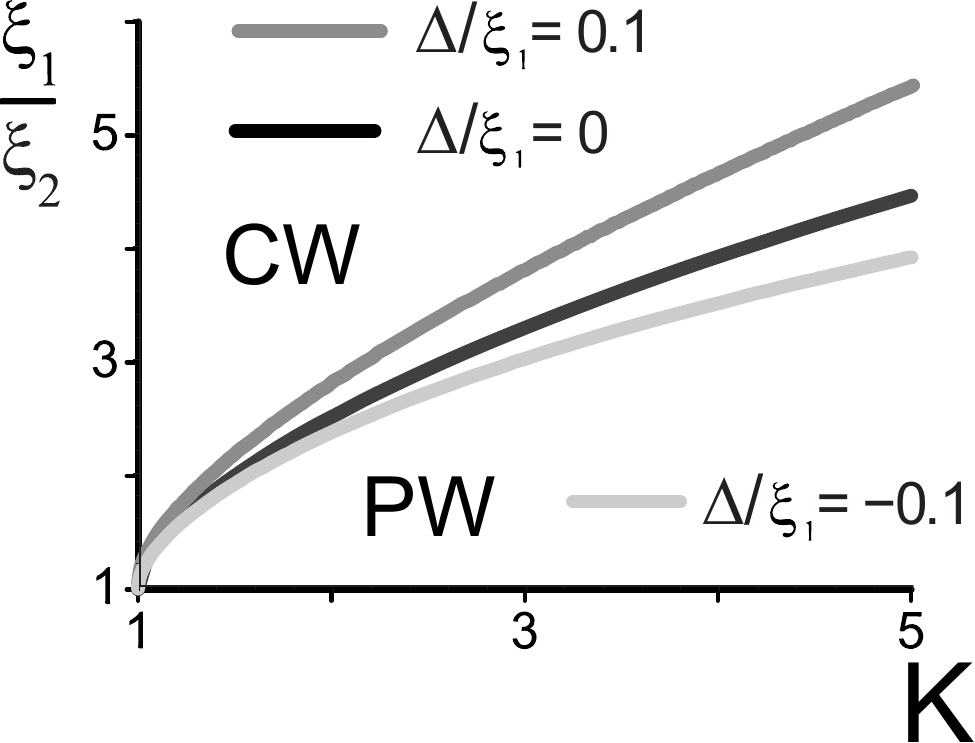}
         \caption{Wetting phase diagram at bulk two-phase coexistence for soft
         walls, in the plane of relative interaction strength
         $K$ and $\xi_1/\xi_2$ for
         different values of the relative trap
        displacement $\Delta$, defined
         in expression~\eqref{gap}. For
            $\Delta=0$, we reproduce the first-order wetting
            line, found earlier in the wetting phase diagram of Fig.~\ref{fig1}.
            Displacement of species $2$ closer to the
            surface, implied by $\Delta < 0$, makes it possible that the wetting transition turns to critical wetting (which it does at least for $K \approx 2$). Furthermore, 
            the parameter region in which complete wetting occurs, broadens. On           the other hand, for
            $\Delta > 0$, the partial-wetting region broadens and the first-order character of the wetting transition is found to persist (at least for $K \approx 2$).  
       \label{fig18}
        }
\end{center}
\end{figure}

\begin{figure}
\begin{center}
	\includegraphics[width=0.22\textwidth]{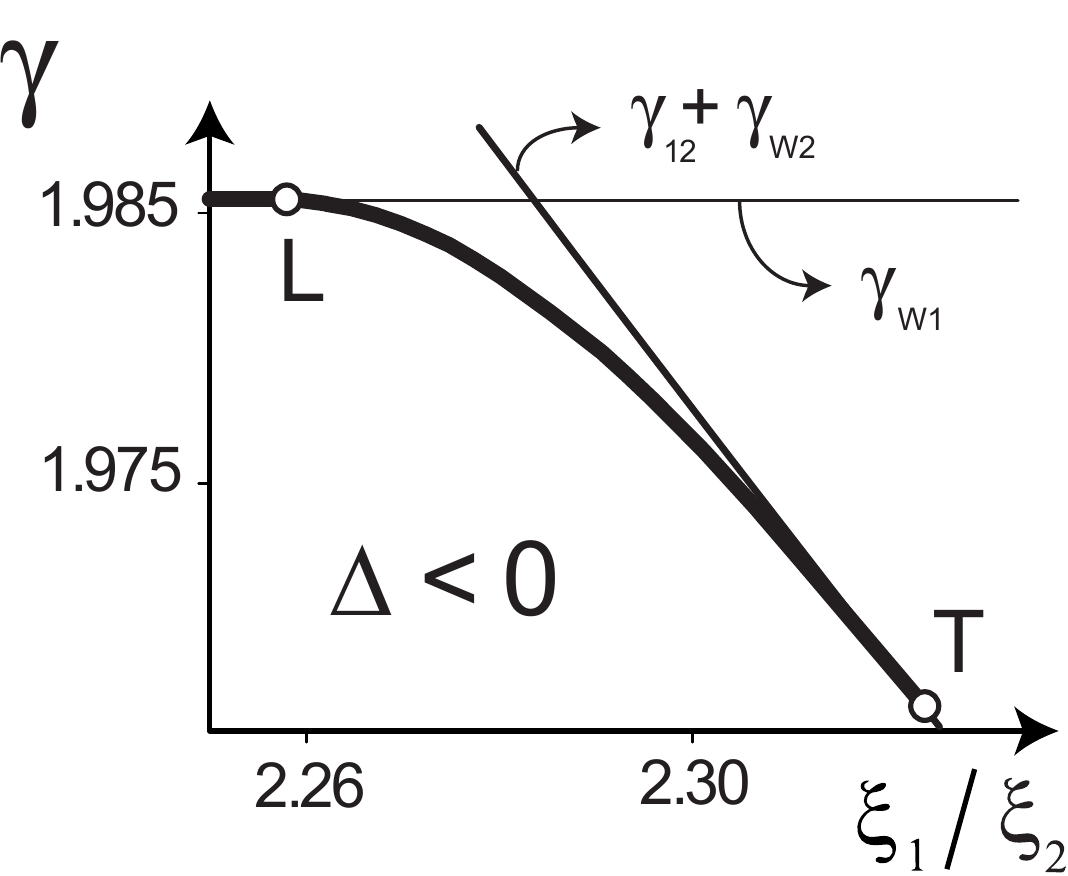}
	\includegraphics[width=0.22\textwidth]{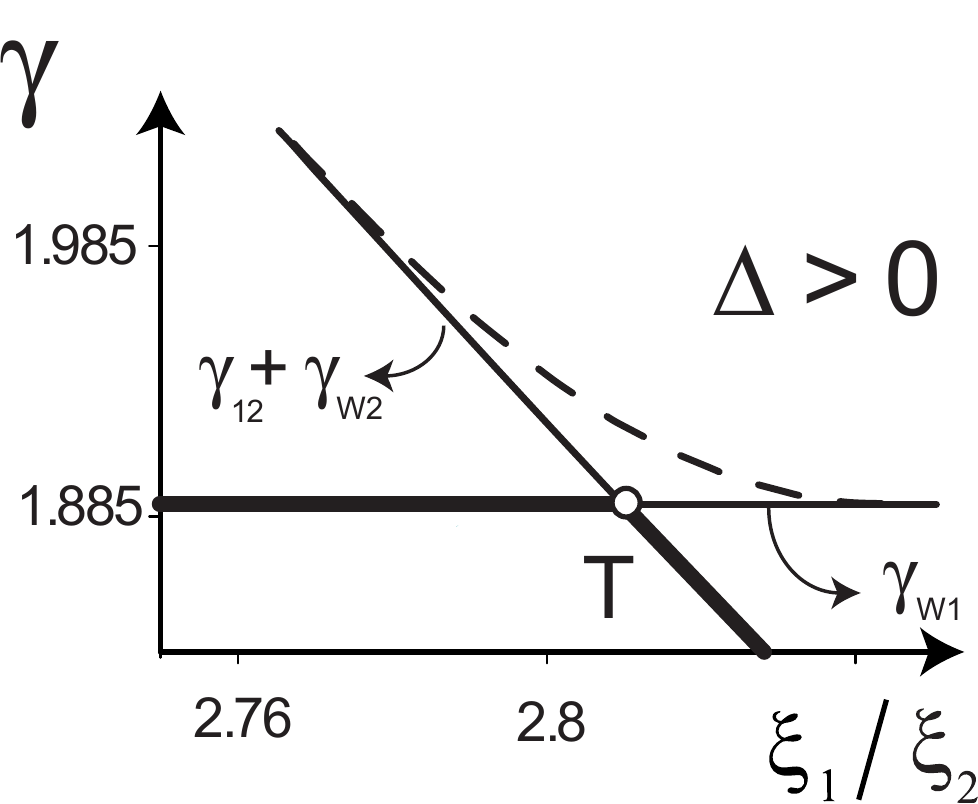}
            \caption{The excess grand potential per unit area
            $\gamma$, at bulk two-phase coexistence, in units of $\xi_1P_1$ against the ratio $\xi_1/\xi_2$
            for $K=2$ for relative trap displacement $\Delta=-0.1\xi_{1}$ (left) and $\Delta=0.1\xi_{1}$
            (right). Note that we vary $\xi_2$ while keeping $\xi_1$ fixed. (Left) A critical transition to a complete
            wetting state takes place at point $T$. Point $L$ is the critical nucleation point for the thin film of species 2. (Right) For $\Delta=0.1\xi_{1}$, a first-order transition to
            complete wetting occurs at point $T$.
       \label{fig17}
        }
\end{center}
\end{figure}

These considerations are combined with numerical analysis in order
to obtain the phase diagram for wetting at soft walls at bulk two-phase
coexistence. In Fig.~\ref{fig18} we draw the wetting phase
boundaries for three values of the relative trap displacement. If
the relative trap displacement is set to zero, i.e., $\Delta=0$, we
recover the case of wetting at a hard wall as studied in
Sect.~\ref{ssec_diagr} where we found that the wetting line is
described exactly by the relation~\eqref{nucleatiecond}~\footnote{Note
that the diagram of Fig.~\ref{fig18} is drawn as a function of
$\xi_1/\xi_2$ and $K$ which are the reciprocal variables of the
ones used in Fig.~\ref{fig1}.}. This first-order phase boundary 
separates the complete wetting regime (CW) from the partial
wetting (PW) regime. For $\Delta=-0.1\xi_{1}$, i.e.,~the
wall of species $2$ is shifted to the left (i.e., into the half space $z<0$), species $2$
tends to be more favored by the wall, which is reflected in the
wetting diagram by an enlargement of the complete wetting region. The
phase boundary (light gray line) which marks the 
wetting transition is, at least in part, critical. Indeed, we have found that a critical wetting transition is possible, whereas for the hard-wall case only first-order wetting occurs. To illustrate the nature of the
transition occuring for $\Delta=-0.1\xi_{1}$ and fixed $K=2$, we plot in
Fig.~\ref{fig17} (left) the excess grand potential $\gamma$ as a
function of  $\xi_1/\xi_2$. It is conspicuous that a critical wetting transition takes place at or very near point $T$. 

\begin{figure}
\begin{center}
	\includegraphics[width=0.45\textwidth]{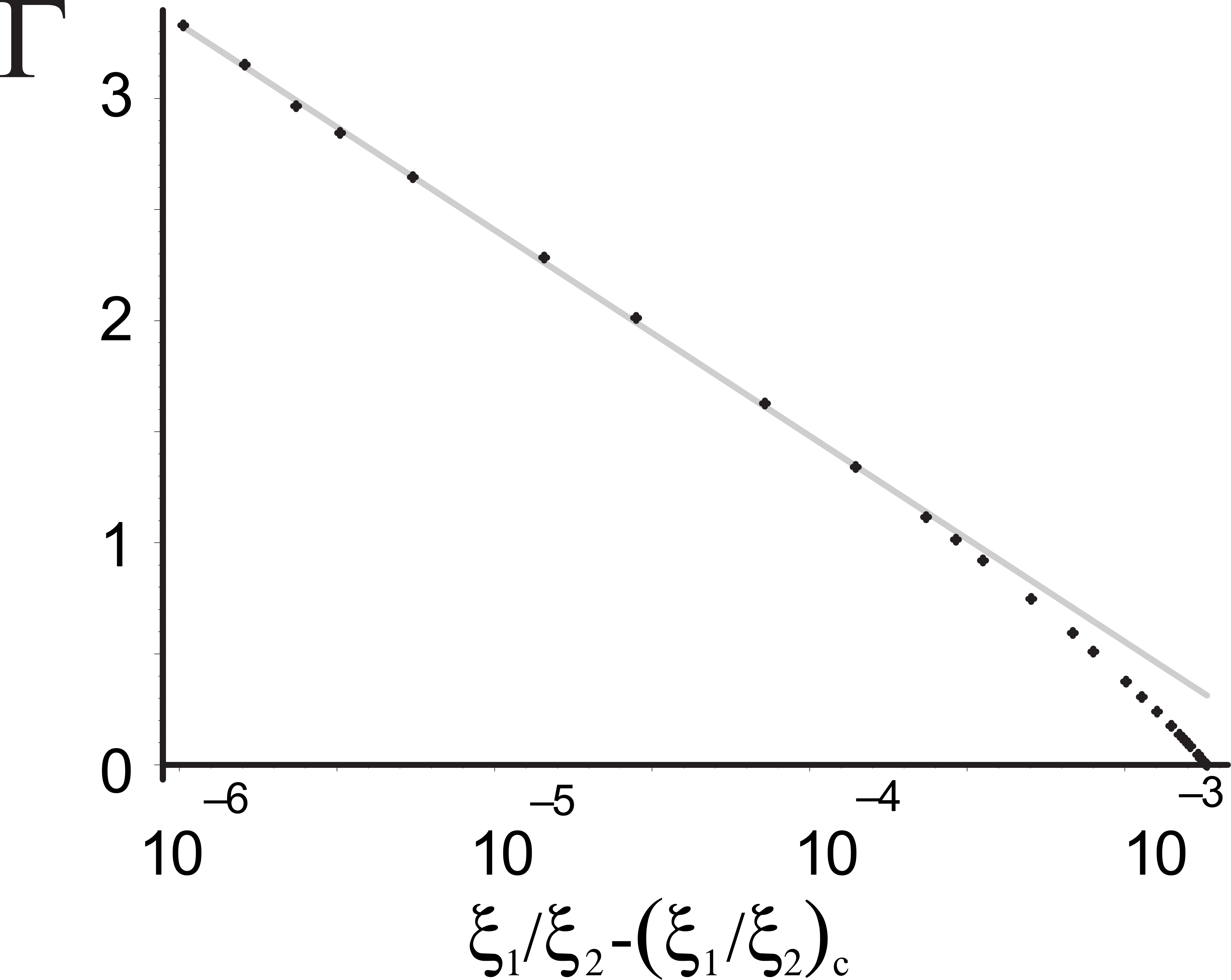}
         \caption{The adsorption of species 2, $\Gamma$ (see Eq.~\eqref{adsorption}), versus the surface field ``distance" to the wetting transition $\xi_1/\xi_2 - (\xi_1/\xi_2)_c$, on a semi-log plot. The critical wetting point consistent with the apparent logarithmic divergence of the adsorption is located at $(\xi_1/\xi_2)_c=2.3227$ (point $T$ in Fig.~\ref{fig17} (left)). The line is a linear fit to the leftmost ten points on the curve.
       \label{fig33}
        }
\end{center}
\end{figure}

Note that a purely numerical
analysis is insufficient for proving the existence of a continuous wetting transition. The transition might still be very weakly first-order. However, the behavior of the adsorption, defined in~\eqref{adsorption}, displaying a {\em logarithmic divergence} approaching the wetting transition point $T$, is strongly indicative of critical wetting (for a system with short-range forces). Fig.~\ref{fig33} shows the computed adsorption of species 2 upon approach of the wetting transition.

Consider now
the wetting transition for $\Delta=0.1\xi_{1}$ as indicated
in Fig.~\ref{fig18} by the upper (medium gray) line. This wetting phase boundary, or at least a part of it,
is of {\em first order}. This is exemplified by Fig.~\ref{fig17} (right), in which the discontinuity of the 
slope of the ground-state excess grand potential per unit area
$\gamma$ is conspicuous, when the system goes over to a complete wetting state at point $T$.

We may conclude that in case  the two shifted
hard walls coincide, i.e.,~when $\Delta=0$,
one recovers (to a good approximation, with an error of order $[\lambda_i/\xi_i]^5$ in the spreading coefficient) the extraordinary wetting
scenario of Sect.~\ref{ssec_diagr} found for a hard wall. The associated fist-order wetting transition may turn into a critical one for $\Delta<0$, whereas a first-order transition may persist for $\Delta>0$. We have not studied the precise extent of the regions of first-order and critical wetting in the phase diagram. We have not investigated the order of the critical wetting transition, nor have we studied the possibility of tricritical wetting or other phenomena that might be present. We come back to these issues in the Conclusion and Outlook section.

\section{Experimental Relevance}\label{sec_exper}
The two ingredients that are essential for the
experimental realization of our set-up are the
ability to adjust the wetting parameters, which depend on the atomic constants, and the ``hard wall''
potential. We now argue that both are accessible and can be manipulated in
state-of-the-art experiments.

Ultracold atomic gases possess the
exceptional feature that both the sign and the strength of the
interactions can be altered~\cite{inouye,papp,kasevich,
rychtarik}. A dramatic variation in the scattering
length is observed near the matching of the energy of two free
atoms with the energy of their bound state. The matching can be
performed since an externally applied magnetic field induces
different Zeeman shifts for the bound state on the one hand and
for the free atoms on the other hand. Recent observation of these
\textit{Feshbach resonances} in multi-component
systems~\cite{papp} proved the ability of independently regulating
any one of the three present scattering lengths $a_{11}$, $a_{22}$
and $a_{12}$. For the experimental exploration of our surface
system, we suggest a tuning of the interspecies scattering length
$a_{12}$. This is adequate, firstly, because $a_{12}$ linearly probes the parameter $K$; second, because varying the parameters $a_{11}$ and $a_{22}$ would influence both $K$ and $\overline{\xi}_2/\xi_1$ (see expression~\eqref{scatter}); and third, because variation of $a_{12}$ limits the loss of atoms in the condensate (by three-body collisions)  to the interfacial zone.

The hard walls introduced here are more than just a textbook
example; by means of blue-detuned evanescent wave atomic mirrors,
current experiments are able to produce steep
walls~\cite{kasevich,rychtarik,savalli,landragin,marani,perrin,kaiser,westbrook, bender}. The
evanescent electromagnetic wave is entailed at the surface of a
dielectric prism from total internal reflection of a linearly
polarized blue-detuned laser beam. The ``blue detuning" means that
the externally applied frequency is higher than an atomic
resonance frequency which causes the induced dipole to be
out of phase with the applied signal. If the amplitude of the
potential caused by the evanescent wave is sufficiently high
($\text{max}_{z}(U(z))>\mu_i$), the atoms are not attracted by the
van der Waals potential very close to the prism but feel a
repulsive barrier which has the form:
\begin{align}\label{surfacepotential1}
U(z)=U_{0}e^{-z/\lambda}.
\end{align}
The amplitude $U_0$ is proportional to the inverse frequency detuning from resonance \cite{westbrook}. The decay length $\lambda$ of the potential is chiefly determined
by the wavelength of reflected light and is as small 
as 50 nm in several systems of experimental  interest (see further). 
This length must be compared with the healing length of
the BEC which typically is in the range from $200$ to $400$ nm, but as argued before, it can
be tuned by a Feshbach resonance. Finally, to confine the atoms
near to the wall, one may use a conventional harmonic trap for
$z>0$ which needs to be sufficiently flat-bottomed at the
center.

The relative trap displacement $\Delta$, defined in~\eqref{gap}, depends on various physical parameters and it is not evident how it can be varied experimentally, and whether it can be varied independently of varying the surface field $\xi_1/\xi_2 -1$ or the interaction strength $K$. To shed some light on this, we consider the case of a mixture of two species consisting of the same atoms and the same isotopes, but different hyperfine states. For such mixtures, assuming a single wavelength (single laser) generating the evanescent wave emanating from a prism, and {\em provided} the detuning of the laser frequency from the atomic resonance frequency is large compared to the frequency corresponding to the hyperfine splitting, we can simplify our discussion and consider wall potentials characterized by
\begin{subequations}
\begin{align}
\lambda_1&=\lambda_2 \equiv \lambda,
\\
U_{10}& = U_{20} \equiv U_0.\label{symplify}
\end{align}
\end{subequations}
Since $\mu_i \propto \xi_i^{-2}$, we obtain, using~\eqref{lengths}, the simple expression
\begin{align}\label{simpleDelta}
\Delta = \Delta_2-\Delta_1 \approx 3.205 \lambda\, (\,[\lambda/\xi_1]^{2}- [\lambda/\xi_2]^{2}),
\end{align}
to leading order in $\lambda/\xi_i$.
From this result we learn that the relative trap displacement is largely controlled by the individual healing lengths $\xi_i$ and that it is possible to leave $\Delta$ as well as the surface field $\xi_1/\xi_2 -1$ unchanged, when we vary $K$ by manipulating only the mixed scattering length $a_{12}$, by making use of the Feshbach resonance technique, for example (see Eq.~\eqref{scatter}). 

Another important piece of information provided by the result~\eqref{simpleDelta}, concerns the sign of $\Delta$. It is conspicuous, since we assume that phase 2 is preferentially absorbed at the wall ($\xi_2 < \xi_1$), that $\Delta$ is a negative quantity. This implies that for BEC mixtures of the same isotope adsorbed at an optical wall {\em critical wetting} is a possibility (cf. the wetting phase diagram discussed in the previous section).

We now ask whether we can predict the value of $\Delta$ reliably for experimentally relevant systems. Since our main result~\eqref{lengths} is a truncated expansion in the ratio $\lambda_i/\xi_i$, our approach is only meaningful for systems in which this ratio is significantly smaller than unity. In several experimental systems this condition is fulfilled, indeed. Calculating $\lambda$ using the well-known expression for the decay constant of the evanescent wave, given in, e.g., \cite{landragin}, we retrieve
\begin{align}
\lambda=\frac{\lambda_L}{4\pi\sqrt{n^2\sin^2\phi-1}},\label{decayEW}
\end{align}
where $\lambda_L$ is the laser wavelength, $n$ the index of refraction of the surface coating of the prism and $\phi$ the angle of incidence. This leads to the values $\lambda = 43$ nm \cite{kasevich}, $47$ nm \cite{perrin}, $48$ nm \cite{kaiser}, $ 56$ nm \cite{landragin,savalli,marani}    and $70$ nm \cite{rychtarik}. On the other hand, the healing length is typically 200  to 400 nm. For example, for $^{23}$Na 200 nm has been reported~\cite{ketterle_review} and 400 nm has been reported for the widely used $^{87}$Rb~\cite{leggett}. This means that for the experimental systems considered, our supposedly small parameter $\lambda/\xi$ ranges from $0.1$ to $0.3$. Therefore, if we compare the correction term of order $[\lambda/\xi]^2$ in~\eqref{lengths} to the preceding constant (1.154), we obtain a relative correction of 3$\%$ to 25$\%$. This signifies that, for some systems of experimental relevance, our expansion in $\lambda/\xi$ is useful. Finally, we recall that in order to obtain a quantitative estimate for $\Delta$ it is also important to take into account that the amplitudes $U_{0i}$ depend sensitively on the frequency detunings, which may differ considerably for the different species, even for the same isotopes (unless the detuning is large compared to the hyperfine splitting). Note that the detunings implemented experimentally vary from fractions of a GHz to about 100 GHz~\cite{kaiser, kasevich, savalli, landragin, marani}.

\section{Wetting in a Trap}\label{sec_trap}
\begin{figure}
\begin{center}
\includegraphics[width=0.45\textwidth]{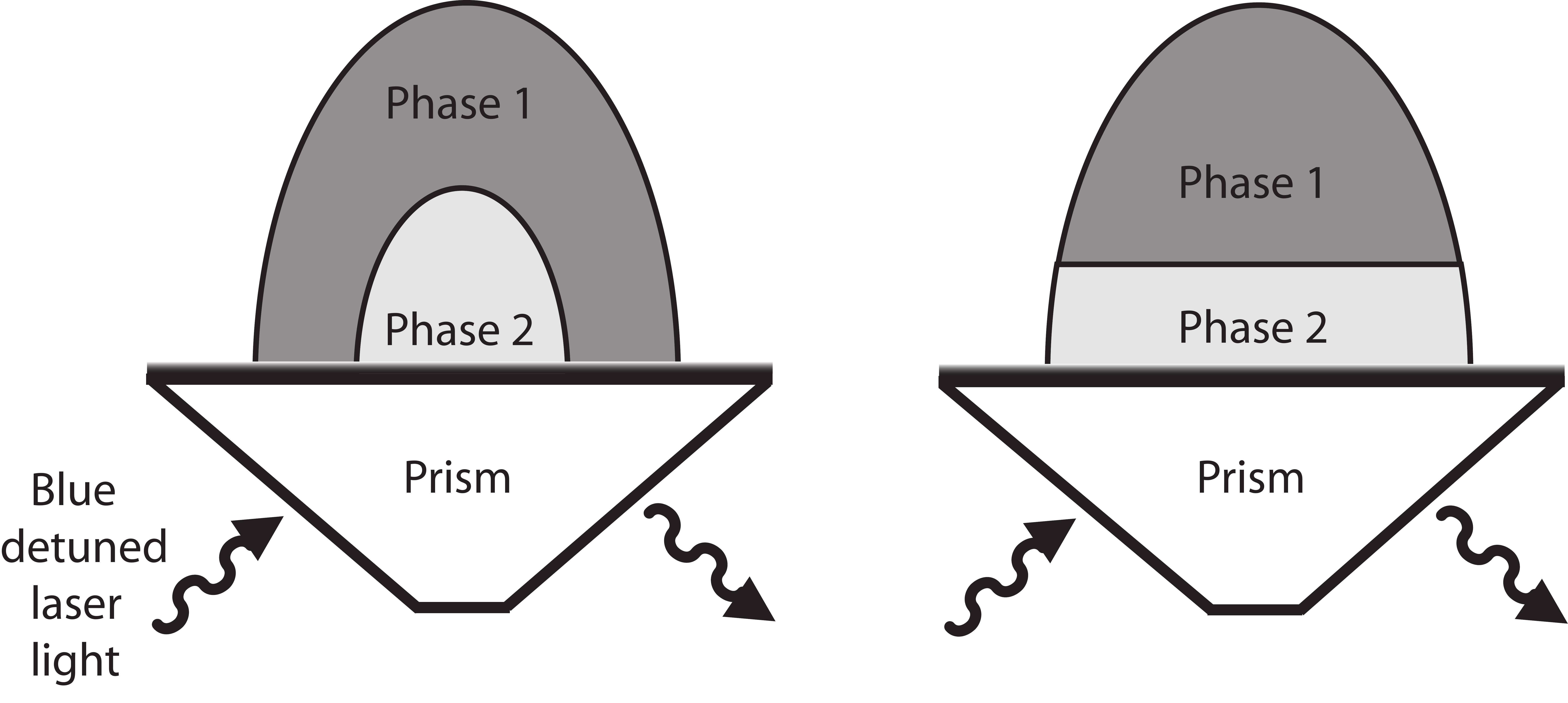}
        \caption{Possible experimental set-up for observing the wetting
transition in a trap: Atoms are contained by
        a (anisotropic) harmonic confinement and are held up by an
evanescent wave prism. The figures show cross sections of partial
wetting (left)
        and complete wetting (right) configurations for a trapped
binary mixture at bulk two-phase coexistence. As argued in Sect.~\ref{sec_trap}, the
        wetting characteristics do not depend on the position along a
hard wall.\label{fig19}}
   \end{center}
\end{figure}

Assume now that, as proposed before, we contain a binary BEC in a
harmonic trap and we introduce a hard surface which cuts the
trap in two (see Fig.~\ref{fig19}). A natural question is then whether the
wetting characteristics vary along the hard wall when the species are at
two-phase coexistence along the wall. The answer is negative: Since both the
condition~\eqref{voorwaarde} and the wave functions
$\widetilde{\psi}_i$ only depend on the variables $K$ and
$\overline{\xi}_2/\xi_1$ and these can be expressed in terms of the
scattering lengths and the masses alone (see Eqs.~\eqref{scatter}),
the wetting properties do not depend on the position. 

The underlying assumption here is that the
characteristic harmonic oscillator length $L$ associated with the
(harmonic) magnetic trap $U_{harm}$ is large compared 
to the
healing length (see condition~(4)). This implies that locally,
at position $\mathbf{r}$, the effective chemical potential $\mu_i$ can
be replaced with a \textit{local} chemical potential
$\mu_i-U_{harm}(\mathbf{r})$  (see also the discussion in Section III of
the first article of Ref.~\cite{BVS1}). This is akin to the local density
approximation to the chemical potential (Section 12.5 in~\cite{pitaevskii}).

On the other hand, when working with a soft wall potential, one must
look at expression~\eqref{CWcondition2} which is the condition for
complete wetting. This condition, as well as the wave functions
themselves depend on the parameter $\Delta/\xi_1$, the position
dependence of which goes as
$\Delta/\xi_1\propto\sqrt{\mu_1-U_{harm}(\mathbf{r})}$ for a harmonic
trap. Thus, whereas the wetting properties at a hard wall are
position-independent, they may be position-dependent for a soft wall.

\section{Conclusion and outlook}\label{sec_concl}

In this paper we have predicted, based on the Gross-Pitaevskii theory for binary mixtures (species 1 and 2) of Bose-Einstein condensates at $T=0$, that wetting and prewetting phase transitions are possible when the mixture is adsorbed at an optical wall. The optical wall consists of an evanescent wave emanating from a prism in which laser light is totally internally reflected. The wall is represented by an external potential acting on the condensates, which turns on exponentially with a decay length $\lambda$ that is small compared to the healing length $\xi$ of a condensate. We have revisited the limit $\lambda \rightarrow  0$ (hard wall) for which wetting and prewetting transitions were predicted in our foregoing Letter \cite{indekeu}. For this limit we provide the exact expressions for the first-order wetting phase boundary and for the prewetting surface, in the global phase diagram. 

Our main results pertain to the more realistic softer wall, at finite $\lambda/\xi$. We provide useful expansions in this ratio, for all relevant surface excess quantities. We find that, provided that we may truncate the expansions at order $[\lambda/\xi]^4$, the soft-wall problem can be captured by introducing two hard walls, one for each condensate, shifted in space by an amount that can be calculated perturbatively, and that depends in a simple manner on the soft-wall parameters and the condensate healing lengths.

In the hard-wall limit the wetting transition is of first order \cite{indekeu}. We have demonstrated that for soft walls, this needs no longer be so.
Indeed, for a range of soft wall parameters, we have found that the wetting transition is critical. It is characterized by a continuous, logarithmic, divergence of the wetting layer thickness and by a continuous first derivative of the spreading coefficient at wetting. In particular, for an adsorbed BEC mixture consisting of identical isotopes but different hyperfine states, which is a case of great experimental relevance, {\em critical wetting} is possible. We have also established that in other regions of the parameter space, the wetting transition is of first order. We have illustrated the experimental usefulness of our computations by identifying several cases for which our expansion variable $\lambda/\xi$ is indeed smaller than unity. We argue that for exploring the wetting phase diagram experimentally, it would suffice to manipulate primarily the mixed scattering length $a_{12}$, which can be done with the Feshbach resonance technique. This parameter directly affects the strength of the interspecies atomic repulsion.

What remains to be investigated is the precise extent of the regions of first-order and critical wetting in the global phase diagram for soft walls. In particular, the separatrices between the first-order and critical regimes have to be identified. Is the cross-over governed by a tricritical wetting transition, or is there a critical endpoint scenario \cite{BonnRMP}? More interestingly still, does the global phase diagram feature infinite-order wetting transitions and/or {\em non-universal} critical wetting, for which the critical exponent of the spreading coefficient depends on the ratio of two lengths \cite{Koga}? For our system these two lengths are the decay lengths of the order parameters towards their bulk values in the wetting phase, which is pure phase 2 in our set-up. Consequently, the relevant lengths are, on the one hand, the healing length $\xi_2$ of condensate 2, and, on the other hand, the penetration depth $\xi_1/\sqrt{K-1}$ of condensate 1 (into condensate 2). The critical exponents at wetting may depend continuously on the ratio of these two lengths. Alternatively, it is also possible that the critical wetting transition is the {\em universal} second-order wetting transition. These scenarios are not mutually exclusive. Both possibilities can be realized. These fascinating questions will be the subject of future research on this problem.

\section{Acknowledgements}
J.O.I. thanks Marek Napi\'orkowski for a stimulating discussion motivating Eq.~\eqref{overlapping}. We thank Michael Goldman for constructive remarks on the manuscript and gratefully acknowledge KU Leuven Research Grant No.~OT/11/063.

\appendix
\section{Nucleation: analytic solution}\label{sec_appendix1}
\begin{figure}[h]
		\includegraphics[width=0.45\textwidth]{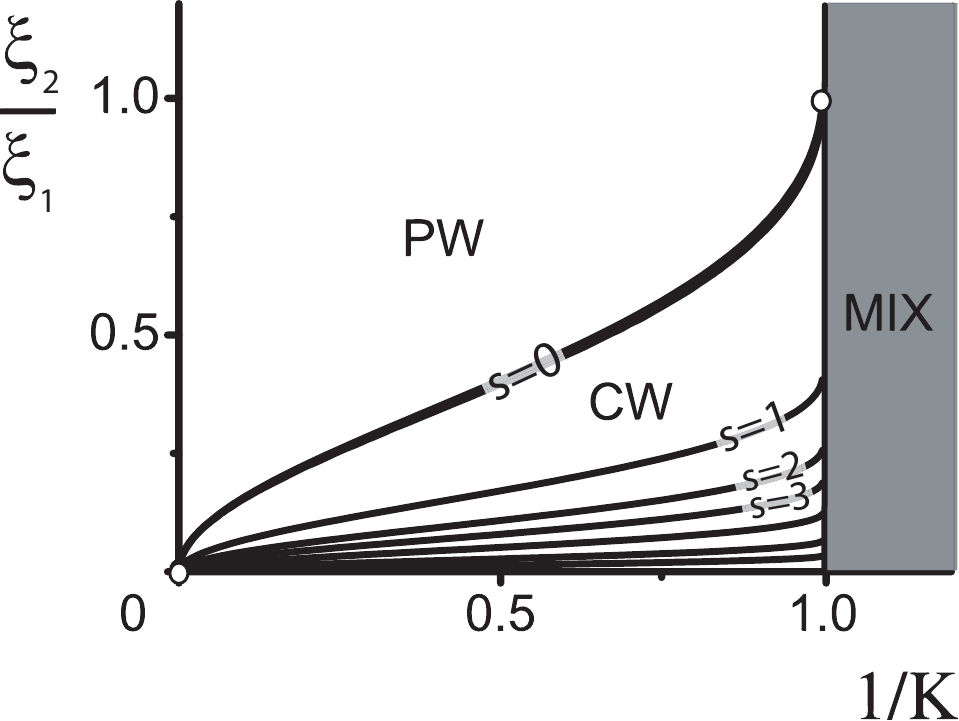}
       \caption{
        Wetting phase diagram in the plane of inverse relative interaction parameter
$1/K$ and surface-field-related parameter $\xi_{2}/\xi_{1}$. The lines
numbered with the integer $s$ correspond to the solutions~\eqref{nucleated} and indicate the loci where excitations
(with $s$ nodal planes) of zero energy exist \cite{BVSPhD}. Below each of these
lines these excitations have a negative energy. The $s=0$ line
is the wetting phase boundary, which coincides precisely with the nucleation line for infinitesimal films of phase 2.
The lines with $s=1,2,\ldots$ lie in the complete wetting (CW)
regime and have no physical significance, since the
equilibrium state in this regime is a macroscopic wetting layer of species $2$.
        \label{patr_fig1}}
\end{figure}

Here we prove that ~\eqref{oplossing} solves
Eq.~\eqref{nlvgl1} together with the boundary
conditions~\eqref{boundcond}, only when relation
~\eqref{nucleatiecond} is satisfied. The problem is equivalent to the Schr\"odinger bound state problem in a potential $V(\tilde z) = - 1/\cosh ^2 (\tilde z/\sqrt{2}) $. The solution to Eq.~\eqref{nlvgl1} that remains finite for $\tilde z \rightarrow \infty$
is~\cite{landau}:
\begin{align}\label{nucleationsolution}
&\widetilde{\psi}_2(\widetilde{z})=\\
&\frac{F \left[
A^{+},A^{-};(A^{+}+A^{-}+1)/2;(1-\tanh
\left(\widetilde{z}/\sqrt{2})\right)/2
 \right ]}{\cosh\left(\widetilde{z}/\sqrt{2}\right)^{\sqrt{2}[\xi_2/\xi_1]^{-1}\sqrt{K-1}}},\nonumber
\end{align}
with $F$ the hypergeometric function and
\begin{align}\label{constanten}
A^{\pm}&=\frac{1}{2}+\frac{\sqrt
{2(K-1)}}{[\xi_2/\xi_1]}\pm\frac{\sqrt {1+8
\,[\xi_2/\xi_1]^{-2}K}}{2}.
\end{align}
The Dirichlet boundary condition yields~\cite{abramowitz}:
\begin{align}\label{nucleationzero}
\widetilde{\psi}_2(0)=\frac{\sqrt{\pi}\,\Gamma\left[(A^{+}+A^{-}+1)/2\right]}{\Gamma\left[(A^{+}+1)/2\right]
\Gamma\left[(A^{-}+1)/2\right]}=0,
 \end{align}
with $\Gamma$ the gamma function. Since $A^{+}>0$, the only possibilities for satisfying the boundary condition are given by
$A^{-}+1=-2s$ for $s=0,\,1,\ldots$ so that:
\begin{align}\label{nucleated}
\frac{\sqrt{K-1}}{[\xi_2/\xi_1]}=\frac{\sqrt{2}([\xi_2/\xi_1]^{-2}-1-3s-2s^2)}{3+4s}.
\end{align}
The integer $s$ counts the number of nodes of the nucleated wave
function $\widetilde{\psi}_2$ and one can check that for $s>0$,
all curves in the $\xi_2/\xi_1-K$ plane which are determined by
relation~\eqref{nucleated}, lie in the CW region of
Fig.~\ref{fig1}. This is shown in Fig.~\ref{patr_fig1}. Therefore, for physical nucleation only the solution with $s=0$ is relevant, which yields expressions~\eqref{nucleatiecond}
and~\eqref{oplossing}. Note that recently analogous solutions were found in Ref.~\cite{pikhitsa} representing bound solutions of a single BEC within an identical geometrical set-up.

\section{Derivation of the surface tension for the soft wall}\label{sec_softwall}
Starting from the system with one single condensate near a hard
wall, we can continuously soften the wall by turning a
\textit{confining length} $\lambda_{_i}$ to a nonzero value when
we assume the surface potential to be of the form:
\begin{align}\label{surfacepotential2}
U_{_i}(z)=U_{_{i0}}e^{-z/\lambda_{_i}}.
\end{align}
We calculate in the following the resulting excess energy per unit
area. First we rescale the $z$-coordinate to
$-\widehat{z}\xi_{_i}=-z+\lambda_{_i}\ln [U_{_{i0}}/\mu_{_i}]$ so
as to get for the GP Eq.~\eqref{GP2}:
\begin{align}\label{gpsurfacepotential}
\ddot{\widetilde{\psi}}_{_i}=\widetilde{\psi}_{_i}\left(-1+e^{-\widehat{z}/[\lambda_{_i}/\xi_{_i}]}+\widetilde{\psi}_{_i}^2\right).
\end{align}
We change the coordinate $\widehat{z}$ to the variable
$\chi_{_i}$, which we define as:
\begin{align*}
\chi_{_i}=[\lambda_{_i}/\xi_{_i}]^2e^{-\widehat{z}/[\lambda_{_i}/\xi_{_i}]}.
\end{align*}
Again, as was the case for the $1$-$2$ interface at strong
segregation, the relaxation of $\lambda_{_i}$ introduces two
effects: firstly, the $\tanh$ profile will shift and
secondly, the wave function will be distorted over the length
$\lambda_{_i}/\xi_{_i}$. To separate the part of the wave function
$\widetilde{\psi}_{_i}$ which is modified due to the surface
potential from the shifted $\tanh$ profile, we rewrite
$\widetilde{\psi}_{_i}$ in the form:
\begin{align*}
\widetilde{\psi}_{_i}=\breve{\psi}^{_0}_{_i}+[\lambda_{_i}/\xi_{_i}](\breve{\psi}_{_{i0}}-\breve{\psi}_{_{i0}}^{_0})
+[\lambda_{_i}/\xi_{_i}]^3(\breve{\psi}_{_{i1}}-\breve{\psi}_{_{i1}}^{_0})+\ldots,
\end{align*}
After a rescaling of the spatial coordinate
$\widehat{z}\equiv\breve{z}[\lambda_{_i}/\xi_{_i}]$, we can define
$\breve{\psi}^{_{i0}}$, $\breve{\psi}^{_0}_{_{i0}}$ and
$\breve{\psi}^{_0}_{_{i1}}$ by:
\begin{align*}
\breve{\psi}^{_0}_{_i}&=\Theta(\breve{z}+\delta_{_0}
+[\lambda_{_i}/\xi_{_i}]^2\delta_{_1})\\
&\quad\times\tanh\left[\frac{[\lambda_{_i}/\xi_{_i}](\breve{z}+\delta_{_0}
+[\lambda_{_i}/\xi_{_i}]^2\delta_{_1})}{\sqrt{2}}\right]\\
&\equiv [\lambda_{_i}/\xi_{_i}]\breve{\psi}_{_{i0}}^{_0}
+[\lambda_{_i}/\xi_{_i}]^3\breve{\psi}_{_{i1}}^{_0}+\ldots
\end{align*}
We substitute all in the GP Eq.~\eqref{gpsurfacepotential}
which to first and second order yields:
\begin{subequations}
\begin{align*}
\dot{\breve{\psi}}_{_{i0}}+\chi_{_i}\ddot{\breve{\psi}}_{_{i0}}&=\breve{\psi}_{_{i0}},\\
\chi_{_i}\dot{\breve{\psi}}_{_{i1}}+\chi^2_{_i}\ddot{\breve{\psi}}_{_{i1}}&=-\breve{\psi}_{_{i0}}+\chi_{_i}
\breve{\psi}_{_{i1}},\label{second}
\end{align*}
\end{subequations}
where the overdot denotes the derivative with respect to
$\chi_{_i}$. Note that, as opposed to the equations found in
Refs.~\cite{lundh,dalfovo,khawaja}, we arrive at a linear equation,
the reason for which lies in the boundary conditions. The
solution for $\breve{\psi}_{_{i0}}$ is:
\begin{align*}
\breve{\psi}_{_{i0}}=\sqrt{2}K_{_0}(2\sqrt{\chi_{_i}})
\end{align*}
with $K_{_0}$ the modified Bessel function of the second kind and
when $\breve{z}\rightarrow\infty$, we find that
$\breve{\psi}_{_{i0}}(\breve{z})=(\breve{z}+\delta_{_0})/\sqrt{2}$
with $\delta_{_0}=-2\left(\ln[\lambda_{_i}/\xi_{_i}]+A\right)$
where $A=0.577\ldots$, the Euler-Mascheroni constant. By a
numerical calculation of $\breve{\psi}_{_{i1}}$, we also found
that $\delta_{_1}=3.205\ldots$ One can then expand the surface
tension~\cite{BVSPhD,BVS1} as:
\begin{align*}
\gamma_{_{Wi}}=\gamma_{_{i0}}+\gamma_{_{i1}}[\lambda_{_i}/\xi_{_i}]+\gamma_{_{i3}}[\lambda_{_i}/\xi_{_i}]^3+\gamma_{_{i5}}[\lambda_{_i}/\xi_{_i}]^5+\ldots
\end{align*}
  and calculations lead to the result:
\begin{align*}
\gamma_{_{i0}}&=\frac{4\sqrt{2}}{3}P_{_i}\xi_{_i},\\
\gamma_{_{i1}}&=P_{_i}\xi_{_i}\left[\ln\left(
[U_{_{i0}}/\mu_{_i}][\lambda_{_i}/\xi_{_i}]^2\right)+1.154\right],\\
\gamma_{_{i3}}&=-3.205P_{_i}\xi_{_i}\\
  \gamma_{_{i5}}&=\lim_{L\rightarrow
\infty}-P_{_i}\xi_{_i}\int_{_{-\infty}}^{_L}
\breve{\psi}_{_{i0}}^4\text{d}\breve{z}+P_{_i}\xi_{_i}\int_{_{-\delta_{_0}}}^{_L}
\frac{(\breve{z}+\delta_{_0})^4}{4}\text{d}\breve{z}\\
&=-12.028\, P_{_i}\xi_{_i}.
\end{align*}
The first-order and third-order terms in $\lambda_{_i}/\xi_{_i}$
result from a shift of the $\tanh$ profile and the fifth-order
term arises from distortions of the wave function. In fact the
fifth-order term has an additional contribution arising from the
fifth-order shift of the $\tanh$ which can be calculated by
introducing a higher-order correction to the wave function
$\breve{\psi}_{_{i2}}$.

\end{document}